\documentclass[twocolumn,english,aps,prd,showpacs,floatfix,superscriptaddress,
notitlepage,longbibliography]{revtex4-1}
\usepackage[T1]{fontenc}
\usepackage[latin9]{inputenc}
\usepackage{amsmath}

\makeatletter

\usepackage{microtype}

\usepackage{dcolumn}
\usepackage{graphicx}
\usepackage[small,bf,hang,raggedright,FIGTOPCAP]{subfigure}
\usepackage{varioref}
\usepackage{epstopdf}
\usepackage{color}
\usepackage{eucal}
\usepackage{float}
\usepackage{soul}
\graphicspath{{./pictures/}}
\usepackage{braket}

\usepackage{blkarray}

\DeclareMathOperator{\Tr}{Tr}

\makeatletter

\DeclareRobustCommand{\cev}[1]{%
  \mathpalette\do@cev{#1}%
}
\newcommand{\do@cev}[2]{%
  \fix@cev{#1}{+}%
  \reflectbox{$\m@th#1\vec{\reflectbox{$\fix@cev{#1}{-}\m@th#1#2\fix@cev{#1}{+}$}}$}%
  \fix@cev{#1}{-}%
}
\newcommand{\fix@cev}[2]{%
  \ifx#1\displaystyle
    \mkern#23mu
  \else
    \ifx#1\textstyle
      \mkern#23mu
    \else
      \ifx#1\scriptstyle
        \mkern#22mu
      \else
        \mkern#22mu
      \fi
    \fi
  \fi
}

\makeatother

\usepackage{babel}
\begin{document}

\title{Interband tunneling effects on materials transport properties using the first principles Wigner distribution}

\date{\today}

\author{Andrea Cepellotti}
\email{acepellotti@g.harvard.edu}
\author{Boris Kozinsky}
\affiliation{Harvard John A. Paulson School of Engineering and Applied Sciences, Harvard University, Cambridge, MA 02138, USA
}

\begin{abstract}
Electronic transport in narrow gap semiconductors is characterized by spontaneous vertical transitions between carriers in the valence and conduction bands, a phenomenon also known as Zener tunneling.
However, this effect is not captured by existing models based on the Boltzmann transport equation.
In this work, we propose a new fully first principles model for electronic transport using the Wigner distribution function and implement it to solve the equations of motion for electrons.
The formalism generalizes the Boltzmann equation to materials with strong interband coupling and include transport contributions from off-diagonal components of the charge current operator.
We illustrate the method with a study of Bi$_2$Se$_3$, showing that interband tunneling dominates the electron transport dynamics at experimentally relevant small doping concentrations, a behavior that is likely shared with other semiconductors, including topological insulators.
Surprisingly, Zener tunneling occurs also between band subvalleys separated by energy much larger than the band gap.
\end{abstract}

\maketitle

\section{Introduction}

Small band gap semiconductors are expected to possess transport characteristics that differ significantly from conventional wide-gap materials due to Zener (or Klein) tunneling, a phenomenon in which carriers undergo vertical transitions across the band gap, which can substantially increase electrical conductivity.
This phenomenon is relevant for a number of novel systems, such as small band gap nanotubes or multi-layer graphene-based systems \cite{PhysRevB.82.045416,zener_1934}. 
Topological insulators are another important example, thanks to their variety of interesting physical properties and promising applications, such as low-power electronics and quantum computing \cite{Yan_2012, RevModPhys.90.015001, doi:10.1002/pssr.201206411, Heremans2017, Liu2019}.
Topological insulators are characterized by conductive surfaces, while their bulk phases have a small band gap, and they may therefore display Zener tunneling and enhanced bulk conduction, which warrants investigation.

%%%%%%%%%%%%%%%%%%%%%%%%%%%%%%%%%%%%%%%%%%%%%%%%%%%%%%%%

The de-facto tool of choice for first-principles simulation studies of electronic transport properties in crystals is the ab initio Boltzmann Transport Equation (aiBTE), which provides estimates of transport properties in good agreement with experimental measurements (e.g. \cite{PhysRevB.83.205208, PhysRevB.92.075405, Ponce:2016, PhysRevB.97.121201, PhysRevB.94.085204, doi:10.1002/aenm.201870095}). 
However, this semiclassical approach is not always sufficient to model electronic transport properties. Zener tunneling, for example, is not captured by semiclassical models, as they omit contributions from off-diagonal components of the flux operators.
Sophisticated models based on the non-equilibrium Green's function methods \cite{reviewNEGF} are capable of overcoming the limitations of semiclassical models.
However, lack of efficient numerical and theoretical formulations prevents the combination of these approaches with first-principles computations that can treat realistic materials, and neglect the critical influence of scattering on the charge flux.
Here we explore a new approach based on the Wigner function: this formalism is particularly appealing as it is exact in principle and, as discussed below, reduces to the aiBTE when off-diagonal flux components are neglected.
The use of the Wigner function to study materials' electronic properties has been limited so far to studies of model Hamiltonians (see e.g. Ref. \cite{doi:10.1063/1.5046663} for a recent review).
In contrast, first-principles modeling can provide not only a more accurate description of real materials' Hamiltonian, but also an accurate description of electronic scattering that is responsible for the relaxation of the out-of-equilibrium system.
This work formulates the generalized transport approach building on recent progress in first-principles computations of electron scattering rates and fills this methodological gap, opening possibility for the study of first-principles electronic properties of realistic materials.

%%%%%%%%%%%%%%%%%%%%%%%%%%%%%%%%%%%%%%%%%%%%%%%%%%%%%%%%
In this work, we derive and implement from first-principles an equation of motion for electrons, termed the ab initio Wigner transport equation (aiWTE) based on the single-particle Wigner distribution function.
The formalism is capable of describing the space-time evolution of electrons, including effects due to electronic scattering, and captures the off-diagonal contributions of flux operators to transport properties.
This equation is explicitly solved for the set of electronic transport coefficients, in particular electrical conductivity and Seebeck coefficient, showing how they can be significantly different from their semiclassical counterpart due to the presence of off-diagonal components of the Wigner distribution function.
We apply the formalism to a first-principles study of the bulk transport properties of the topological insulator Bi$_2$Se$_3$.
We show that at small doping concentrations, the aiWTE estimates of bulk electronic transport properties deviate significantly from semiclassical estimates, due to the presence of Zener tunneling, i.e. vertical transitions that couple carriers of valence and conduction bands.
%%%%%%%%%%%%%%%%%%%%%%%%%%%%%%%%%%%%%%%%%%%%%%%%%%%%%%%%%%%%%%%%%%%%%%

\section{Theory}

We start by considering the ground state Hamiltonian $H_0$ of a crystal, which we assume to be an independent-particle Hamiltonian with eigenvalues $\epsilon_{bs}(\boldsymbol{k})$ and Bloch states $\psi_{\boldsymbol{k}bs}(\boldsymbol{x})$, where $\boldsymbol{x}$ is the position, $\boldsymbol{k}$ the wavevector, $b$ the band index and $s=\pm 1$ the spin index.
The ground-state is perturbed by a constant electric-field $\boldsymbol{E}$ which couples with the carriers' charge $e$ and by the electron-phonon interaction $H_{el-ph}$, so that the total Hamiltonian is $H = H_0 + e \boldsymbol{x} \cdot \boldsymbol{E} + H_{el-ph}$.
To derive an equation of motion for such a system, we use the single-particle Wigner function $f$ of the system \cite{wigner1932}, defined as the Wigner transform of the density matrix $\rho$ as 
\begin{equation}
f_{bb'ss'}(\boldsymbol{x},\boldsymbol{k},t) 
= 
\sum_{\Delta \boldsymbol{k}} e^{i \Delta \boldsymbol{k} \cdot\boldsymbol{x}} \rho_{bb'ss'} \Big(\boldsymbol{k}+\frac{\Delta \boldsymbol{k}}{2},\boldsymbol{k}- \frac{\Delta \boldsymbol{k}}{2};t \Big) \;,
\end{equation}
where $t$ is the time. 
We build the Wigner function through a transformation of the density matrix $\rho_{bb'ss'}(\boldsymbol{k},\boldsymbol{k}')$ in the reciprocal space representation.
Such Wigner transform consists in a rotation of variables $\boldsymbol{k},\boldsymbol{k}' \to \frac{\boldsymbol{k}+\boldsymbol{k}'}{2}, \boldsymbol{k}'-\boldsymbol{k}$ combined with a Fourier transform on one variable.
The Wigner function operates in a phase-space representation, which is especially useful to draw connections between quantum and classical mechanics.

The evolution of the Wigner function \cite{moyal1949,Groenewold1946} is found through a Wigner transform of the equation of motion of the density matrix, and has been shown to be
\begin{align}
\frac{\partial f_{bb'ss'}(\boldsymbol{x},\boldsymbol{k},t)}{\partial t} 
=& 
- \big\{ \big\{ f(\boldsymbol{x},\boldsymbol{k},t), H(\boldsymbol{x},\boldsymbol{k}) \big\} \big\}_{bb'ss'} \\
:=&
\frac{i}{\hbar} \Big( f(\boldsymbol{x},\boldsymbol{k},t) \star H(\boldsymbol{x},\boldsymbol{k})  \nonumber \\
& \quad - H(\boldsymbol{x},\boldsymbol{k}) \star f(\boldsymbol{x},\boldsymbol{k},t) \Big)_{bb'ss'} 
\;,
\end{align}
where $\{\{f,H\}\}$ is the Moyal bracket (the quantum mechanical extension of the Poisson bracket) and the Moyal product $\star$ is defined as
\begin{equation}
f \star H
= 
f(\boldsymbol{x},\boldsymbol{k}) 
\exp\bigg(\frac{i}{2} \bigg( \frac{\cev{\partial}}{\partial \boldsymbol{x}} \cdot \frac{\vec{\partial}}{\partial \boldsymbol{k}} - \frac{\cev{\partial}}{\partial \boldsymbol{k}} \cdot \frac{\vec{\partial}}{\partial \boldsymbol{x}} \bigg) \bigg) 
H(\boldsymbol{x},\boldsymbol{k}) \;,
\end{equation}
where $H(\boldsymbol{x},\boldsymbol{k})$ is the Wigner transform of the Hamiltonian, and the arrows indicate that the derivative operator is acting to the left/right operators.
%To ensure that such derivatives exist, we choose a particular wavefunction gauge, as discussed in the supplementary information.

We now simplify the Hamiltonian supposing that the electron-phonon interaction is weak and evaluate the Moyal bracket for the single-particle part of the Hamiltonian; the electron-phonon interaction is added later as a perturbation.
Since we are interested in macroscopic properties, we can make the assumption that only slow spatial variations of the Wigner function are relevant.
Therefore, we expand the Moyal product in Taylor series to the lowest orders of $\hbar$ and find an equation of motion which we term the ab initio Wigner Transport Equation (aiWTE), that is
\begin{widetext}
\begin{align}
\begin{split}
\frac{\partial f_{bb'ss'}(\boldsymbol{x},\boldsymbol{k},t)}{\partial t}
&+
 \frac{i}{\hbar} \Big[ \mathcal{E}(\boldsymbol{k}) + \boldsymbol{d}(\boldsymbol{k})\cdot\boldsymbol{E} , f(\boldsymbol{x},\boldsymbol{k},t) \Big]_{bb'ss'}
+
 \frac{1}{2} \Big\{ \boldsymbol{v}(\boldsymbol{k}) , \cdot \frac{\partial f(\boldsymbol{x},\boldsymbol{k},t)}{\partial \boldsymbol{x}} \Big \}_{bb'ss'} -  \\
&-
 e \boldsymbol{E} \cdot \frac{\partial f_{bb'ss'}(\boldsymbol{x},\boldsymbol{k},t)}{\partial \boldsymbol{k}}
=
-\frac{\partial f_{bb'ss'}(\boldsymbol{x},\boldsymbol{k},t)}{\partial t} \bigg|_{coll} 
\;.
\label{aiWTE}
\end{split}
\end{align}
\end{widetext}
where $\{,\}$ is an anticommutator, $\mathcal{E}_{bb'ss'}(\boldsymbol{k}) = \delta_{bb'}\delta_{ss'} \epsilon_{bs}(\boldsymbol{k})$ is a tensor of electronic energies, $\boldsymbol{d}_{bb'ss'}(\boldsymbol{k}) = (1-\delta_{bb'})\Braket{\boldsymbol{k}bs | e\boldsymbol{x} | \boldsymbol{k}b's'}$ is a tensor of electric dipoles between two Bloch states (typically used to describe optical excitations), and $\boldsymbol{v}_{bb'ss'}(\boldsymbol{k})$ is the velocity operator, defined from the commutator of the Hamiltonian and the position operator as $\boldsymbol{v}=\frac{i}{\hbar}[H,\boldsymbol{r}]$.
The electron-phonon scattering operator $\frac{\partial f_{bb'ss'}(\boldsymbol{x},\boldsymbol{k},t)}{\partial t} \big|_{coll}$ is added as a perturbation to the aiWTE and is built using scattering rates from the Fermi Golden rule \cite{simoncelli2019, PhysRevB.96.115420, Zhan2016, NEDJALKOV2013, Nedjalkov2011}.
We refer to the Supplementary Information for a more detailed derivation of the aiWTE.

The aiWTE needs to be solved to obtain an estimate of the single-particle Wigner distribution function.
As a first comment, we note that the aiBTE is recovered as a limiting case of the aiWTE, when the off-diagonal terms $b\neq b'$ and $s\neq s'$ are set to zero. 
This corresponds physically to a situation when different bands do not couple via Zener tunneling.
This can happen, for example, when neither thermal excitation nor dipole interaction provide sufficient energy to allow for the vertical transition of one particle from one band to another.
Therefore, the most interesting terms to discuss in the aiWTE are the off-diagonal terms, which introduce the possibility of additional electronic transitions, or couplings, between different electronic states at a given wavevector $\boldsymbol{k}$.
We further note that some of the off-diagonal terms of the aiWTE shown here are absent in other works based on either the density matrix or the Wigner function \cite{RevModPhys.74.895, PhysRevB.53.4870, PhysRevB.35.9644, PhysRevB.86.155433, kane_2015, PhysRevB.96.144303}.
Additionally, we note that the electronic aiWTE is conceptually similar to a formalism developed for phonon transport in Ref. \cite{simoncelli2019}, although here we use a simplified derivation and include the effect of external forces (the electric field).

%To better understand the off-diagonal terms in Eq. \ref{aiWTE}, it is illustrative to estimate the electrical conductivity $\sigma$.
The aiWTE can be solved to estimate transport coefficients with a technique similar to the one used for the aiBTE.
For steady-state transport in a bulk system, the Wigner distribution is stationary in time and independent from the particular position inside the bulk; therefore, we look for a solution in the form $f_{bb'ss'}(\boldsymbol{k})$.
Next, we look for the linear response to an applied external electric field $\boldsymbol{E}$ or a temperature gradient $ \boldsymbol{\nabla} T$ and write $ f_{bb'ss'}(\boldsymbol{k}) = \bar{f}_{bs}(\boldsymbol{k}) + f^{E_i}_{bb'ss'}(\boldsymbol{k}) E_i$ or $f_{bb'ss'}(\boldsymbol{k}) = \bar{f}_{bs}(\boldsymbol{k}) + f^{T_i}_{bb'ss'}(\boldsymbol{k}) \nabla_i T$, where $i$ is the direction of the applied perturbations and $\bar{f}_{bs}(\boldsymbol{k})$ is the Fermi--Dirac distribution function.
We further adopt the relaxation time approximation, which simplifies the treatment of the scattering operator.
After inserting these two proposed solutions in the aiWTE, and retaining only terms up to linear order in $\boldsymbol{E}$ or $ \boldsymbol{\nabla} T$, we find two equations for the diagonal part of the aiWTE:
\begin{equation}
e v^i_{bbss}(\boldsymbol{k}) \frac{\partial f_{bbss}(\boldsymbol{k})}{\partial \epsilon} 
=  
 \Gamma_{bs}(\boldsymbol{k}) f^{E_i}_{bbss}(\boldsymbol{k})
 \;.
\end{equation}
and 
\begin{equation}
v^i_{bbss}(\boldsymbol{k}) \cdot \frac{\partial f_{bbss}(\boldsymbol{k})}{\partial T} 
=  
 \Gamma_{bs}(\boldsymbol{k}) f^{T_i}_{bbss}(\boldsymbol{k})
 \;,
\end{equation}
where $\Gamma_{bs}(\boldsymbol{k})$ is the carrier's linewidth.
These equations are equivalent to the standard aiBTE for describing electronic transport within the relaxation time approximation.
However, the aiWTE also gives rise to two equations for the off-diagonal terms of the Wigner distribution, which are
\begin{widetext}
\begin{equation}
i \Big(\epsilon_{bs}(\boldsymbol{k})-\epsilon_{b's'}(\boldsymbol{k}) \Big) f^{E_i}_{bb'ss'}(\boldsymbol{k}) 
+ i \Big( \bar{f}_{b's'}(\boldsymbol{k}) - \bar{f}_{bs}(\boldsymbol{k})\Big) d^i_{bb'ss'}(\boldsymbol{k})
=  
- \frac{\Gamma_{bs}(\boldsymbol{k}) + \Gamma_{b's'}(\boldsymbol{k})}{2} f^{E_i}_{bb'ss'}(\boldsymbol{k})  \;,   \quad b\neq b' \;, s\neq s' \;,
\end{equation}
and
\begin{equation}
i \Big(\epsilon_{bs}(\boldsymbol{k})-\epsilon_{b's'}(\boldsymbol{k}) \Big) f^{T_i}_{bb'ss'}(\boldsymbol{k})  
+ \frac{1}{2} \bigg( \frac{\partial \bar{f}_{b's'}(\boldsymbol{k})}{\partial T} + \frac{\partial \bar{f}_{bs}(\boldsymbol{k})}{\partial T} \bigg) v^i_{bb'ss'}(\boldsymbol{k}) 
=  
- \frac{\Gamma_{bs}(\boldsymbol{k}) + \Gamma_{b's'}(\boldsymbol{k})}{2} f^{T_i}_{bb'ss'}(\boldsymbol{k})\;,  \quad b\neq b' \;, s\neq s' \;,
\end{equation}
\end{widetext}
describing the response of the system to electrical and thermal perturbations, respectively.
Now the aiWTE is in a form that can be readily solved for $f^{E_i}$ and $f^{T_i}$ with some algebra, allowing us to reconstruct the Wigner distribution of a crystal in presence of an electric field or a thermal gradient.

Having found the Wigner distribution, transport properties are readily obtained.
For example, the charge current density is
\begin{equation}
\boldsymbol{j}
=
\frac{e}{2VN_k} \sum_{\boldsymbol{k}bs} \Big\{ \boldsymbol{v}(\boldsymbol{k}), f(\boldsymbol{k}) \Big\}_{bbss} \;,
\end{equation}
where $V$ is the crystal unit cell volume and $N_k$ the number of wavevectors used to integrate the Brillouin zone. 
Additionally, the heat flux is 
\begin{equation}
\boldsymbol{q}
=
\frac{1}{2VN_k} \sum_{\boldsymbol{k}bs} (\epsilon_{bs}(\boldsymbol{k})-\mu) \Big\{ \boldsymbol{v}(\boldsymbol{k}), f(\boldsymbol{k}) \Big\}_{bbss} \;.
\end{equation}
These definitions readily allow us to compute transport coefficients using an approach typical of transport theory.
After inserting the solution to the aiWTE in the definition of $\boldsymbol{j}$ and $\boldsymbol{q}$, it is readily seen that charge current and heat flux can be written in the form $\boldsymbol{j} = L_{EE} \boldsymbol{E} + L_{ET} \boldsymbol{\nabla} T$ and $\boldsymbol{q} = L_{TE} \boldsymbol{E} + L_{TT} \boldsymbol{\nabla} T$, where $L$ denotes the Onsager coefficients.
Transport coefficients can be expressed in terms of these Onsager coefficients: the electrical conductivity $\sigma = L_{EE}$, the Seebeck coefficient $S = - L^{-1}_{EE} L_{ET}$, and the thermal conductivity $k=-(L_{TT} - L_{TE} L_{EE}^{-1} L_{ET})$.

We now inspect the electrical conductivity in more details. 
Since $\boldsymbol{j}$ is linear in $f_{bb'ss'}(\boldsymbol{k})$, the total electrical conductivity $\sigma$ is a sum of diagonal and off-diagonal contributions $\sigma_{ij} = \sigma_{ij}^{aiBTE} + \Delta \sigma_{ij}$, where $i,j$ are cartesian labels.
Here, $\sigma_{ij}^{aiBTE}$ comes from the diagonal terms of the Wigner distribution ($b=b'$ and $s=s'$) and is equal to the estimate from the aiBTE:
\begin{equation}
\sigma^{aiBTE}_{ij} =
\frac{e^2}{V N_k} \sum_{\boldsymbol{k}bs}  \frac{\partial \bar{f}_{bs}(\boldsymbol{k})}{\partial \epsilon}   v^i_{bbss}(\boldsymbol{k}) v^j_{bbss}(\boldsymbol{k}) 
\tau_{bs}(\boldsymbol{k})
\;,
\end{equation}
where $\tau_{bs}(\boldsymbol{k})=\hbar / \Gamma_{bs}(\boldsymbol{k})$ is the carriers' lifetime.
This semiclassical conductivity is corrected by an additional term $\Delta \sigma_{ij}$ that can be found inserting $f^{E_i}_{bb'ss'}(\boldsymbol{k})$ in the definition of $\boldsymbol{j}$ and is
\begin{align}
\label{delta_sigma_eq}
\Delta \sigma_{ij} 
= &
\frac{2e^2}{V N_k}  \sum_{\substack{\boldsymbol{k}bb'ss' \\ bs\neq b's'}}  \frac{\bar{f}_{b's'}(\boldsymbol{k})-\bar{f}_{bs}(\boldsymbol{k})}{\epsilon_{b's'}(\boldsymbol{k})-\epsilon_{bs}(\boldsymbol{k})} \times
 \\
& \times
\frac{ v^i_{bb'ss'}(\boldsymbol{k}) v^{j,*}_{b'bs's}(\boldsymbol{k})
( \Gamma_{bs}(\boldsymbol{k}) + \Gamma_{b's'}(\boldsymbol{k}))}
{ 4(\epsilon_{b's'}(\boldsymbol{k})-\epsilon_{bs}(\boldsymbol{k}))^2
+ 
(\Gamma_{bs}(\boldsymbol{k}) + \Gamma_{b's'}(\boldsymbol{k}))^2 
} \nonumber \;.
\end{align}
%where $\Gamma_{bs}(\boldsymbol{k})$ is the electronic linewidth (here, due to electron-phonon interaction) and $\bar{f}_{bs}(\boldsymbol{k})$ the Fermi--Dirac occupation number.
We can now understand the effects of the off-diagonal corrections.
First, the correction $\Delta \sigma$ is positive (note that $\bar{f}_{bs}(\boldsymbol{k})$ is a monotonic function of $\epsilon_{bs}(\boldsymbol{k})$ ), and therefore the aiWTE will always adjust the semiclassical prediction of conductivity to higher values.
Second, the correction depends on a few quantities: the energy difference between electrons and holes, their linewidths, and velocity.
One scenario where this correction is relevant, for example, is 
%Second, %the expression of the electrical conductivity better illustrates the role of the off-diagonal components of the Wigner distribution function.
whenever the energy difference between an electron and a hole is comparable to their linewidth, so that the two carriers interact.
The strength of such interaction is determined by the velocity matrix element $v_{bb'ss'}(\boldsymbol{k})$, i.e. the matrix element for the optical transition.
In addition, a strong dipole coupling can also occur between states that are far from the conduction or valence band edge, as discussed below.
Whenever this off-diagonal correction is large, the system allows for an additional transport mechanism, known as Zener tunneling, in which electrons propagate by tunneling across the band gap.
In contrast, we stress that Zener tunneling, or similar spontaneous vertical electronic transitions, are entirely omitted in the aiBTE formalism, since the aiBTE doesn't include off-diagonal contributions from the velocity operator.

%%%%%%%%%%%%%%%%%%%%%%%%%%%%%%%%%%%%%%%%%%%%%%%%%%%%%%%%%%%%%%%%%%%%%%%%%%%

\section{Computational Methods}
All quantities appearing in the aiWTE are available from first-principles codes and we can therefore apply this formalism using fully ab-initio parameters. 

We use density functional theory as implemented in the plane-wave software suite Quantum-ESPRESSO \cite{giannozzi2009quantum,giannozzi2017advanced}.
To compute the ground state, we use ultrasoft pseudopotentials from the GBRV library \cite{GARRITY2014446}, with the PBEsol functional.
We use an energy cutoff of 80 Ry, and integrate the Brillouin zone with a $8\times 8\times 8$ mesh of k-points.
We build the trigonal unit cell using experimental estimates of the crystal structure \cite{bi2se3_structure}, i.e. with a lattice parameter of 9.839 \AA, and an angle  $\alpha$ such that $\cos \alpha = 0.91068$.
The Wannier functions are computed using $p$ orbitals on both Bi and Se atoms as initial guesses.

Phonon properties, and the electron-phonon matrix elements are computed with density-functional perturbation theory \cite{baroni_revmodphys} on a coarse grid of $4\times 4\times 4$ q-points.
Electron-phonon matrix elements are subsequently interpolated on a finer grid of $39\times 39\times 39$ Brillouin zone wavevectors using a mixed Wannier and linear interpolation \cite{PhysRevB.94.085204}, while electronic energies and velocities are interpolated using Wannier90 \cite{pizzi_2019}.
Transport properties have been implemented in a custom-made software.
The Dirac-delta ensuring energy conservation during an electron-phonon scattering event has been approximated using an adaptive-smearing scheme \cite{Yates2007}.
Transport properties have been converged with respect to the k-points mesh used to integrate the Brillouin zone.
The scattering operator, detailed in the Supplementary Information, is built considering only intrinsic electron-phonon scattering.
The electron-phonon coupling includes the effect of long-range polar interaction, which contributes significantly to transport coefficients, and is evaluated according to the methodology of Ref. \cite{PhysRevLett.115.176401}.
%We verified that the While the The aiBTE (and thus the diagonal components of the aiWTE) can be solved with the complete scattering matrix, using for example iterative or variational approaches \cite{PhysRevB.83.205208, PhysRevB.92.075405, PhysRevB.94.085204}; details of the scattering matrix are discussed in the Supplementary Information.
%Alternatively, the scattering matrix can be simplified with the relaxation time approximation (RTA), as further discussed in the supplementary information, which replaces the complete scattering matrix with just its diagonal matrix elements.
While the scattering operator may be treated beyond the relaxation time approximation with techniques similar to those of Refs. \cite{PhysRevB.83.205208, PhysRevB.92.075405, PhysRevB.94.085204}, all results of this study are obtained within the relaxation time approximation, as it allows for significantly faster simulations.
We note that results beyond the relaxation time approximation would only affect the diagonal components of the Wigner distribution, but not the off-diagonal components since the action of the scattering operator on the off-diagonal components of the Wigner distribution involve only the carriers' lifetimes/linewidths (see Eq. 18 of Supplementary Information).
%We verified that transport properties are only negligibly impacted by the relaxation time approximation using an exact solver of the aiBTE in a methodology similar to Refs. \cite{PhysRevB.83.205208, PhysRevB.92.075405, PhysRevB.94.085204}; since the RTA is significantly faster from a numerical point of view and that the difference between aiWTE and aiBTE is independent from the usage of the RTA, we have adopted it for all results presented in this work.

\section{Results}

\begin{figure}[htp]
\includegraphics{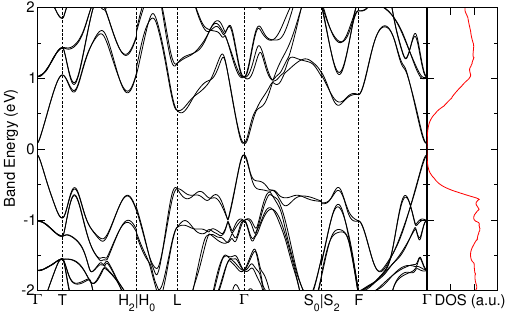}
\caption{
\textbf{Electronic band structure and density of states of Bi$_2$Se$_3$.}
Electronic band structure and density of states (DOS) of Bi$_2$Se$_3$ near the Fermi energy, set at the center of the band gap. 
The bulk crystal is characterized by a small gap, opened by the spin-orbit coupling. 
We also note from the DOS that subvalleys of valence and conduction bands are approximately 1.8 eV apart in energy.}
\label{fig1}
\end{figure}

We now apply the formalism to study the intrinsic phonon-limited electronic transport of bulk Bi$_2$Se$_3$.
In Fig. \ref{fig1} we report the band structure \cite{spglib, HINUMA2017140} and the density of states (DOS) for this narrow-gap semiconductor.
We estimate a quasiparticle gap of 0.2 eV, in agreement with experimental estimates \cite{bi2se3_bandgap}.
We also mention that, while DFT may not always be accurate in estimating the band gap, the qualitative aspects of the results discussed below do not strongly depend on the precise band gap value.
It is worth noting that the DOS increases away from the Fermi level (set at 0 eV at the middle of the band gap) and flattens at energies of approximately -0.8eV and 1.0eV for the valence and conduction bands, respectively, indicating that the subvalleys are separated by an energy of approximately 1.8 eV.

\begin{figure}[ht]
\includegraphics[width=\textwidth]{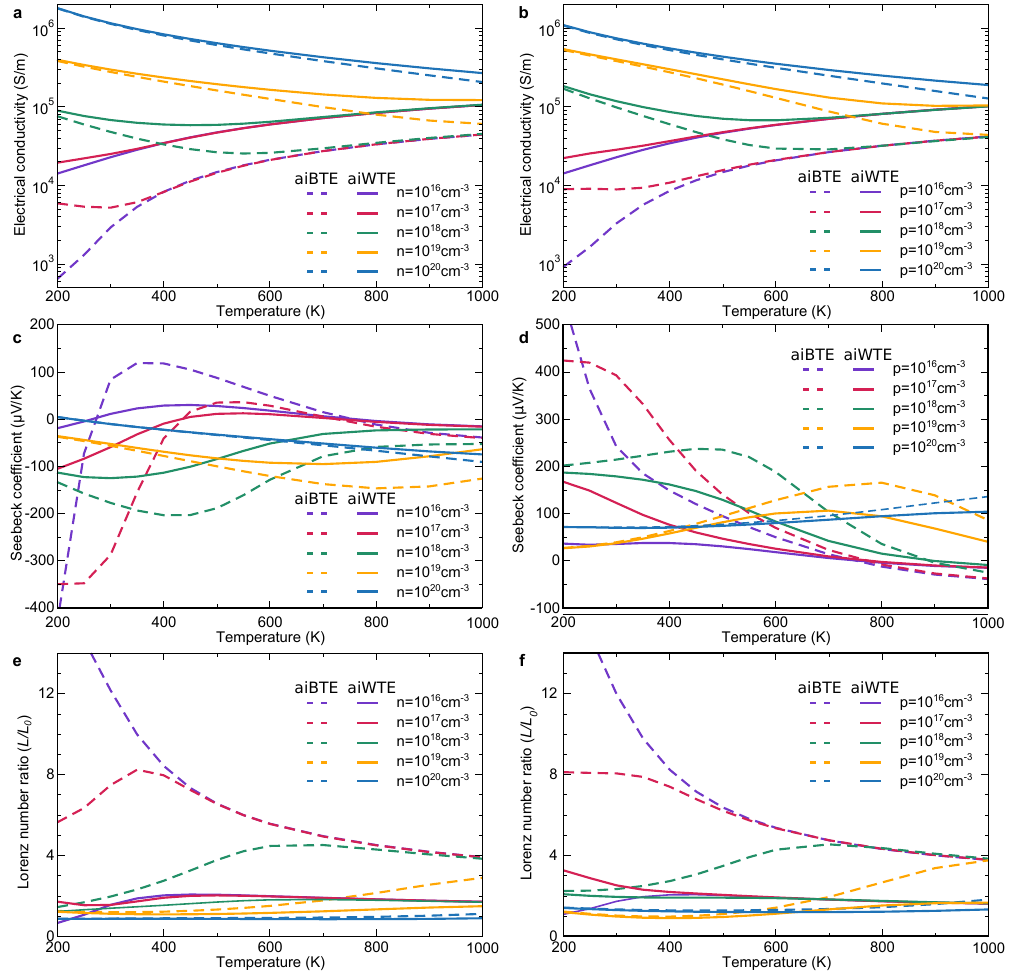}
\caption{
\textbf{Transport properties of Bi$_2$Se$_3$ from semiclassical and Wigner formalism.}
Computational estimates of the in-plane electrical conductivity (panels a and b), Seebeck coefficient (panels c and d), and Lorenz number ratio (panels e and f, see text for description) of Bi$_2$Se$_3$ as a function of temperature, for different values of n-type or p-type doping concentration.
Solid lines are estimated using the Wigner transport equation, while dotted lines are semiclassical estimates obtained solving the Boltzmann transport equation. 
For small doping concentrations, the interaction between electrons and holes significantly affect the estimates of transport coefficients.}
\label{fig2}
\end{figure}

In Fig. \ref{fig2}a (b), we estimate the electron-phonon limited electrical conductivity $\sigma$ of Bi$_2$Se$_3$ in the in-plane direction as a function of temperature for different values of n-type (p-type) doping concentrations. 
Dashed lines represent the semiclassical estimates $\sigma^{aiBTE}$, while aiWTE estimates are shown in solid lines.
For the highest doping values, when the chemical potential shifts into the conduction (valence) band, the conductivity has the typical metallic-like behavior of decreasing with temperature.
Under these conditions, aiBTE and aiWTE do not differ significantly, except at higher temperatures.
In the Supplementary Information, we briefly discuss a comparison with available experimental results which, for the purpose of the present study, shows good qualitative agreement especially at large doping concentrations.

For lower doping concentrations the chemical potential lies in the band gap and we thus observe a semiconducting behavior of $\sigma$ increasing with temperature.
The semiclassical model predicts a smaller conductivity than the aiWTE estimate.
In fact, when only a few carriers from the bottom of the conduction (top of the valence) band are excited, the average carriers' group velocity is small, due to the quadratic nature of the band minimum.
Therefore, the semiclassical contribution to electrical conductivity tends to be rather small.
The aiWTE corrects this picture, including the Zener tunneling effect \cite{zener_1934}.
As carriers from valence and conduction band are close in energy, they can interact and contribute to the electrical transport through the tunneling effect, as discussed above.
For small dopings, the aiWTE correction is significant, and can be much larger than the aiBTE conductivity value.
For the smallest value of doping reported (10$^{16}$cm$^{-3}$), this correction is largest at lower temperatures.
The doping of 10$^{18}$cm$^{-3}$ is an intermediate case, with metallic behavior at low temperatures (and thus smaller aiWTE correction) and semiconducting (with larger aiWTE correction) at higher temperatures as the chemical potential moves from the conduction band into the band gap.
We can thus conclude that a substantial portion of electrical current at low doping is carried through the Zener tunneling included in the aiWTE formalism: the current is not only caused by the carriers traveling at a finite group velocity, but also by carriers' tunneling between single-particle Bloch states.

In Fig. \ref{fig2}, panels c and d, we report the Seebeck coefficient $S$ for n-type (p-type) doping concentrations, with aiBTE results in dashed lines and aiWTE in solid lines.
The Seebeck coefficient is usually expected to have negative values for n-type doping and positive for p-type.
However, there are deviations from this  behavior, and the Seebeck coefficient can even change sign when temperature is varied at fixed doping concentrations.
This complex behavior is not caused by the off-diagonal terms of the Wigner distributions, and the Seebeck coefficient sign changes are observed when using the aiBTE as well, e.g. due to the bipolar effect where minority carriers are thermally excited across the band gap.
Such behavior has been found for Bi$_2$Te$_3$ \cite{https://doi.org/10.1002/aelm.201800904} and CoSb$_3$ \cite{PhysRevB.81.045204}, where the Seebeck coefficient exhibits marked temperature dependence.

The Wigner correction tends to make the Seebeck coefficient smaller in absolute value, which may be attributed to the electrical conductivity appearing at the denominator of the definition of the Seebeck coefficient. As for the case of electrical conductivity, the off-diagonal correction is more prominent at small values of doping concentrations and low temperatures.
This may have a significant consequence when estimating thermoelectric properties, such as the power factor that depends on the square of the Seebeck coefficient.
In this case, the inclusion of off-diagonal components of the Wigner distribution lead to a revision of power factor estimates to smaller values.

%The negative values indicate that a majority of carriers are electrons, although, in a small band gap system, deviations from this behavior can occur.
%In a parabolic band model treated semiclassically, $S$ is  proportional to the logarithmic derivative of the density of states \cite{ziman1960electrons}.
%Therefore, at low temperatures, one expects the Seebeck coefficient to increase as the doping concentration is decreased.
%This phenomenon is crucial to optimize the thermoelectric efficiency, where the goal is to maximize the power factor $\sigma S^2$.
%However, within the aiWTE, $S$ is not anymore simply related to the density of states, as additional terms in the transport equations appear (as detailed in the supplementary information).
%As a result, the large increase of the Seebeck coefficient expected by a semiclassical model at low doping is strongly suppressed by the aiWTE, and $S$ becomes comparable to its values at high doping.
%We thus conclude that the tunneling effects captured by the aiWTE can alter considerably the predictions of thermoelectric properties in narrow-gap semiconductors.

We now examine the relationship of electrical conductivity $\sigma$ and open-circuit electronic thermal conductivity $\kappa_{\text{el}}$. 
The Wiedemann-Franz (WF) law defines the Lorenz number $L=\frac{\kappa_{\text{el}}}{\sigma T}$, which in the ideal metallic limit is a constant $L_0=2.44 \cdot 10^{-8}$ W$\Omega$K$^{-2}$.
Knowledge of $L$ is necessary to decouple the electronic contribution $\kappa_{\text{el}}$ and the lattice contribution $\kappa_{\text{ph}}$ from measurements of the total $\kappa$. 
In Figure \ref{fig2} e and f we plot the computed ratio $\frac{L}{L_0}$ for several temperatures and n-type or p-type kind of doping concentrations, using both aiBTE and aiWTE in dashed and solid lines respectively. 
At high doping the system has metallic character, and both predictions closely follow the WF law. 
In the case of small doping, in the bipolar regime, the semiclassical aiBTE predicts large deviations from the WF law, as has been discussed previously \cite{doi:10.1002/aenm.201870095, doi:10.1063/1.4908244, PhysRevB.85.205410, PhysRevB.99.020305}. 
Remarkably, in the aiWTE solution the Lorenz numbers are much closer to the expected range for semiconductors, indicating that quantum transport effects included in the aiWTE strongly suppress deviations and work towards restoring the validity of the WF law.

\begin{figure}[htp]
\includegraphics{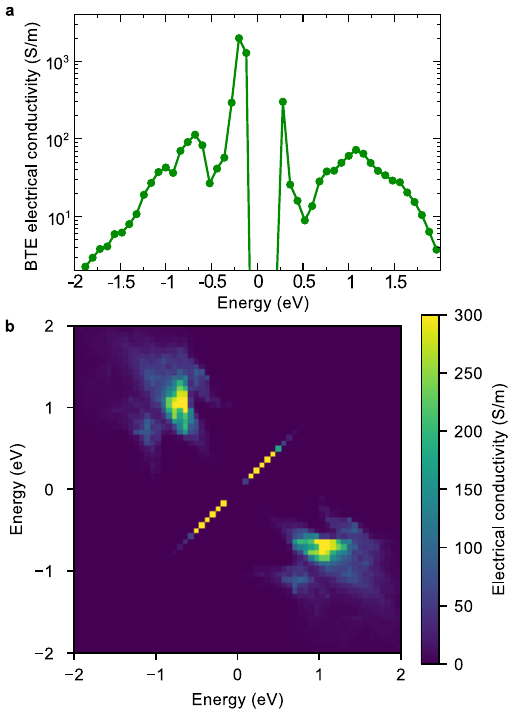}
\caption{
\textbf{Microscopic contributions to electrical conductivity from Boltzmann and Wigner formalisms.}
Panel a: histogram of contributions to semiclassical electrical conductivity as a function of the carrier energy, for Bi$_2$Se$_3$ at 700K and n-doping at 10$^{18}$ cm$^{-3}$. Panel b, 2D histogram of contributions to the electrical conductivity for the same system as estimates with the Wigner transport equation against the energy of two coupled carriers. Off-diagonal contributions represent electrical conductivity arising from the coupling between electrons and holes.}
\label{fig3}
\end{figure}

%We now analyze the energy of the carriers contributing to transport.
In Fig. \ref{fig3}a, we analyze the contributions to the aiBTE electrical conductivity as a function of the carriers' energy at doping concentration of 10$^{18}$ cm$^{-3}$, and temperature of 700 K.
This histogram is built such that the area under the curve integrates to the total electrical conductivity.
Within the semiclassical relaxation time approximation, the quantity plotted is an energy-resolved histogram of $\frac{e^2}{V N_k} \frac{\partial \bar{f}_{bs}(\boldsymbol{k})}{\partial \epsilon} v^2_{bbss}(\boldsymbol{k}) \tau_{bs}(\boldsymbol{k})$, i.e. the contribution of a single mode to the aiBTE electrical conductivity.
As expected, the dominant contributions to electrical conductivity come from carriers whose energy is close to the chemical potential (set at 0 eV).
The contributions of other carriers decay exponentially as their energy gets further from the chemical potential.

The aiWTE correction $\Delta \sigma$ cannot be resolved in terms of a single carrier's energy, since it involves the tunneling between two states at different energies.
Therefore, in Fig. \ref{fig3}b, we plot the contributions to the electrical conductivity as a function of two interacting carriers energies.
On the diagonal, we find again the aiBTE-like terms shown in Fig. \ref{fig3}b.
In addition, we can see important off-diagonal contributions to the electrical conductivity that are not present in the aiBTE and are introduced with the aiWTE.
In particular, there are two peaks of contributions to electrical conductivity, that couple electrons of energy 1.0 eV with holes at -0.8 eV.
These two values correspond to the average energies of the  valence and conduction bands, when the DOS reaches the corresponding maximum values.
Therefore, in contrast to the typical intuition of the Zener tunneling, we find that the most significant coupling between carriers takes place far from the chemical potential, with carriers of energy much larger than the thermal energy.
For this material, the dipole interaction between carriers in subvalleys of the valence and conduction is thus particularly strong, allowing for high-energy carriers to contribute to transport.
As a result, we speculate that Zener tunneling may take place also in semiconductors with a wide gap and contribute significantly to electrical conductivity, provided that the inter-band dipole interaction is sufficiently strong, for instance in materials with high optical absorption character.

\begin{figure}
\includegraphics{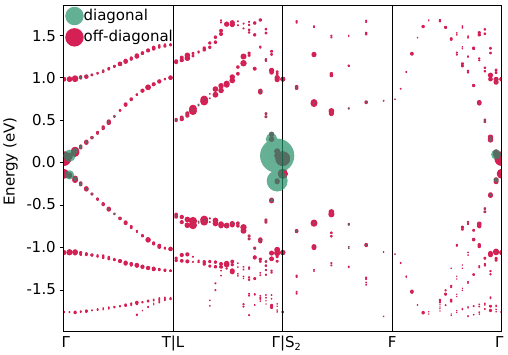}
\caption{
\textbf{Microscopic contributions to electrical conductivity from Boltzmann and Wigner formalisms.}
The dispersion relation of carriers on some high-symmetry lines are plotted together with the contributions to the in-plane electrical conductivity of Bi$_2$Se$_3$ at 700K and doping concentration of $n=10^{18}$ cm$^{-3}$.
In particular, the radius of green dots (mostly visible only near the band gap at the $\Gamma$ point) is proportional to the contribution to electrical conductivity from the diagonal components of the Wigner distribution, while red dots radius is proportional to the contributions of the off-diagonal Wigner distribution.
While semiclassical transport mostly originates close to the band gap at $\Gamma$, the corrections from the aiWTE are coming from several states especially in the conduction and valence subvalleys.}
\label{fig4}
\end{figure}

We can gain further insight by looking at how different points in the Brillouin zone contribute to the conductivity.
In Fig. \ref{fig4}, we plot the contributions to the in-plane electrical conductivity from carriers with wavevector lying on high-symmetry lines of the Brillouin zone, the y-coordinate of the dot in figure is positioned according to the carrier's energy.
The green dots' radius is proportional to the contribution of such carrier to the aiBTE electrical conductivity, i.e. the contributions from the diagonal part of the Wigner distribution, whereas the red dots radius is proportional to the electrical conductivity contribution from the off-diagonal components of the Wigner distribution.
Quantities are computed for the electrical conductivity at temperature of 700K and doping concentration $n=10^{18}$ cm$^{-3}$.
As one would expect from semiclassical arguments, the diagonal contributions are mostly originating from carriers of energy close to the band gap. 
The contributions from the off-diagonal components instead extend to carriers that are further from the band gap edge.
States close to the band gap at the $\Gamma$ point have a sizeable contribution to the off-diagonal conductivity. 
However, subvalleys overall contribute more significantly to the conductivity, especially due to the larger availability of states for the vertical transitions to take place.
We further note that there is not a single transition that dominates the off-diagonal effects.
%We note that diagonal and off-diagonal contributions are not drawn to scale in respect to one another because the radius of diagonal contributions would outsize that of off-diagonal ones: the total area of green dots should equate to $\sigma^{aiBTE}=2.9 \cdot 10^4$ S/m and the red area to $\delta \sigma=4.4 \cdot 10^4$ S/m (the green radius is rescaled by a factor 40 compared to the red radius).
Lastly, we stress that diagonal and off-diagonal contributions have not been drawn to scale with respect to one another.
In fact, diagonal contributions are mostly determined by carriers close to the chemical potential and sum up to a conductivity of $\sigma^{aiBTE}=2.9 \cdot 10^4$ S/m. 
The off-diagonal conductivity $\Delta \sigma=4.4 \cdot 10^4$ S/m receives contributions from carriers over a much larger energy range.
As a result, if the diagonal contributions were drawn to the same scale of the off-diagonal ones, green circles would appear much larger than what is shown; therefore, green radii have been shrunk by a factor about 40 in order to fit in the figure.

\begin{figure}
\includegraphics{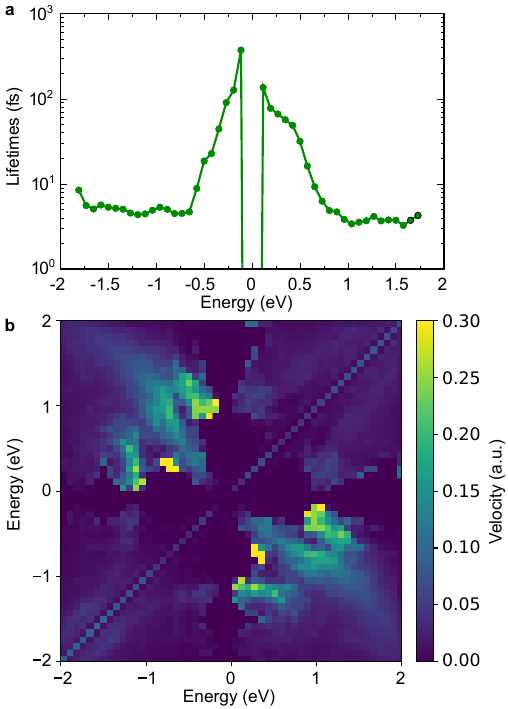}
\caption{
\textbf{Analysis of electronic lifetime and velocity.}
Panel a: average of electronic lifetimes $\tau_{bs}(\boldsymbol{k})$ as a function of energy, for Bi$_2$Se$_3$ at 700K and n-type doping concentration of 10$^{18}$ cm$^{-3}$.
The energy dependence of lifetimes is approximately proportional to the inverse of the density of states.
Panel b: 2D histogram of the velocity operator $v_{bb'ss'}(\boldsymbol{k})$ as a function of the energies of the two states $\epsilon_{bs}(\boldsymbol{k})$ and $\epsilon_{b's'}(\boldsymbol{k})$. 
Diagonal matrix elements are the electronic group velocity, while off-diagonal elements are related to the dipole operator.
The presence of states with large off-diagonal velocity allows the coupling needed for vertical transitions.
}
\label{fig5}
\end{figure}

So far, we have seen that the off-diagonal contributions to electrical conductivity at energies far from the chemical potential arise from a large presence of available states.
However, Eq. \ref{delta_sigma_eq} depends on several other factors.
In Fig. \ref{fig5}a, we plot the average lifetime as a function of carrier's energy, at the same doping density and temperature of Figure \ref{fig3}.
Here we can see that lifetimes tend to be large for states close to the band edges and decrease away from them: to a first approximation, the average lifetime as a function of energy scales inversely proportional to the density of states.
Therefore, the decrease of lifetimes is offset by the larger availability of states, which still allows for a sizeable off-diagonal contributions from states away of the band gap.
In Fig. \ref{fig5}b we also plot the energy-resolved histogram of the velocity operator $|v_{bb'ss'}(\boldsymbol{k})|$ in the in-plane direction, where the two axis represent the energies of the carriers $\epsilon_{bs}(\boldsymbol{k})$ and $\epsilon_{b's'}(\boldsymbol{k})$ respectively, and the color intensity represents the absolute value of the velocity operator matrix element.
Clearly, in order to have a sizeable Wigner correction, it is essential for the velocity operator elements to be sufficiently large.
However, one can see that the energy values where the velocity elements are largest do not always coincide with the energies of states that contribute most to the conductivity.
Therefore, the off-diagonal effects of the Wigner distribution are not easily interpreted in terms of a single factor appearing in Eq. \ref{delta_sigma_eq}: the overall correction arises due to a combination of large linewidth and velocity (dipoles), as well as high availability of states for the transitions.

%%%%%%%%%%%%%%%%%%%%%%%%%%%%%%%%%%%%%%%%%%%%%%%%%%%%%%%%%%%%%%%%%%%%%%

\section{Conclusions}

First-principles simulations of electronic transport are often limited to the semiclassical approximation, that excludes contributions to transport coming from non-semiclassical particle dynamics.
In this study, we have shown that the Moyal equation of motion leads to the Wigner transport equation, an equation of motion for the Wigner function that can be solved using first-principles techniques.
The formalism has a broader range of applicability than the semiclassical Boltzmann transport equation, and includes off-diagonal contributions to charge and heat currents.
In particular, the formalism captures contribution to transport coming from spontaneous vertical transitions between carriers' states, that are especially relevant for narrow-gap semiconductors.
Moreover, the semiclassical Boltzmann transport equation is found as a simplified limit to the Wigner transport equation, provided that off-diagonal components are neglected.
The formalism is well-suited to first-principles simulations and has been tested with a study of the transport properties of Bi$_2$Se$_3$.
We have shown that, while at high doping concentrations the Boltzmann equation provides a fairly accurate description of transport, it fails at low doping concentrations.
At low dopings, spontaneous vertical transitions across the band gap, also known as the Zener tunneling effect, contribute significantly to electronic transport, modifying both electrical conductivity and Seebeck coefficient.
Surprisingly, Zener tunneling does not just take place across the states closest to the band gap, but involves states that are significantly further apart in energy, provided that the dipole interaction is sufficiently strong.
As a result, we have extended the range of applicability of ab-initio transport simulations to materials where band tunneling couples carriers and a semiclassical description is no longer adequate.

%%%%%%%%%%%%%%%%%%%%%%%%%%%%%%%%%%%%%%%%%%%%%%%%%%%%%%%%%%%%%%%%%%%%%%%%%%%%%%%%%%%%%

\subsubsection*{Author contributions}
A. C. derived results and performed numerical simulations. A.C. and B. K. conceived the project and wrote the paper.

\subsubsection*{Competing interests}
The authors declare no competing interests.

\subsubsection*{Data availability}
The datasets generated during and/or analyzed during the current study are available from the corresponding author on reasonable request.

\subsubsection*{Acknowledgments}
We acknowledge funding support from the Star-Friedman Fund for Promising Scientific Research, the Harvard Quantum Initiative, the STC Center for Integrated Quantum Materials, NSF Grant No. DMR-1231319, and computational resources at Harvard FAS Research Computing.

% \bibliography{biblio}


%\newpage
%\clearpage

\onecolumngrid
\section*{Supplementary information}

%%%%%%%%%%%%%%%%%%%%%%%%%%%%%%%%%%%%%%%%%%%%%%%%%%%%%%%%%%%%%%%%%%%%%%%

\subsection{Equation of motion of the Wigner function}

In this section, we detail the derivation of the Wigner transport equation discussed in the main text.

We start from the single-particle Hamiltonian $H$ of an electron in a crystal in presence of an electric field, that is
\begin{equation}
H 
%= H_0 + H_E %+ H_{el-ph} 
= H_0 + e \boldsymbol{x} \cdot \boldsymbol{E} %+ H_{el-ph} 
= H_0 + \boldsymbol{d} \cdot \boldsymbol{E} %+ H_{el-ph} 
\;,
\label{hamiltonian_supp}
\end{equation}
where $H_0$ is the Hamiltonian of a crystal in its ground state, $e$ the electronic charge, $\boldsymbol{x}$ the position operator, $\boldsymbol{E}$ the electric field and $\boldsymbol{\hat{d}} = e \boldsymbol{x}$ the dipole operator.
We further make the hypothesis that the electric field can be added as a perturbation, so that the eigenvectors $\ket{\psi_{\boldsymbol{k} b s}}$ of $H$ are approximately the eigenvectors of $H_0$ as well ($\boldsymbol{k}$ is the wavevector, $b$ the band index, and $s$ the spin index are Bloch quantum numbers).
The eigenvalues of $H_0$ are denoted as $\epsilon_{bs}(\boldsymbol{k})$.
We also stress that the single-particle approximation is consistent with the numerical implementation using parameters computed from density-functional theory.

Before proceeding, it is important to choose a wavefunction gauge such that the derivative $\frac{\partial \ket{\psi_{\boldsymbol{k}bs}}}{\partial \boldsymbol{k}}$ exists and is continuous.
To this aim, we recall the maximally localized Wannier functions, defined as:
\begin{equation}
\ket{\boldsymbol{R}n} 
= 
\frac{V}{(2\pi)^3} \int d\boldsymbol{k} e^{-i\boldsymbol{k}\cdot\boldsymbol{R}} \sum_{bs} U_{\boldsymbol{k},bs n} \ket{\psi_{\boldsymbol{k}bs}} 
= 
\frac{V}{(2\pi)^3} \int d\boldsymbol{k} e^{-i\boldsymbol{k}\cdot\boldsymbol{R}} \ket{\tilde{\psi}_{\boldsymbol{k}n}} 
\;,
\label{supp_hamiltonian}
\end{equation}
where $\boldsymbol{R}$ labels a Bravais lattice site, $U_{\boldsymbol{k},bs n}$ is a tensor fixing the wavefunction gauge, and $n$ is an index over the Wannier functions.
The tensor $U_{\boldsymbol{k},bs n}$ is chosen as the one that maximally localizes Wannier functions \cite{RevModPhys.84.1419}, which has also the benefit of making $\ket{\tilde{\psi}_{\boldsymbol{k}n}}$ a smooth wavefunction across different wavevectors (otherwise, wavefunctions at different wavevectors assume random phases).

Next, we briefly recall the definition of the Wigner transform.
Given an operator in the real-space (position) representation $A(\boldsymbol{x},\boldsymbol{x}')$, we can transform it to the phase-space representation through the Wigner transform $W[\cdot]$, defined as
\begin{equation}
W[ A ]_{nn'}(\boldsymbol{x},\boldsymbol{k}) = 
\int d\boldsymbol{x}' e^{- i 2 \boldsymbol{k}\cdot\boldsymbol{x}'} A_{nn'}(\boldsymbol{x}+\boldsymbol{x}',\boldsymbol{x}-\boldsymbol{x}') \;.
\label{weyl_transform}
\end{equation}
Similarly, if we start from an operator in the momentum representation, we can  transform it into the phase-space representation as
\begin{equation}
W[ A ]_{nn'}(\boldsymbol{x},\boldsymbol{k}) = 
\int d\boldsymbol{k}' e^{i 2 \boldsymbol{k}'\cdot\boldsymbol{x}} A_{nn'}(\boldsymbol{k}+\boldsymbol{k}',\boldsymbol{k}-\boldsymbol{k}') \;.
\label{wigner_transform}
\end{equation}

We now want to describe the equation of motion of the system under the Hamiltonian of Eq. \ref{supp_hamiltonian}.
We can build the density matrix of the system using the basis of wavefunctions in the Wannier gauge $\ket{\tilde{\psi}_{\boldsymbol{k}n}}$ introduced above.
We can thus represent the state of the system using the single-particle density matrix operator $\rho$, whose matrix elements are $\rho_{nn'}(\boldsymbol{k},\boldsymbol{k}';t) = \Tr\{ \hat{\rho}(t) c^\dagger_{\boldsymbol{k}',n'} c_{\boldsymbol{k},n} \} \;,
$, where $c^\dagger_{\boldsymbol{k},n}$, and $c_{\boldsymbol{k},n}$ are creation/annihilation operators of an electronic state $\ket{\tilde{\psi}_{\boldsymbol{k}n}}$.
As described in Ref. \cite{simoncelli2019}, one may use the equation of motion of the density matrix, and Wigner transform it to derive an equation of motion for the system.  
Here, we derive a simplified equation of motion for the state of the system using the Wigner function of the crystal \cite{wigner1932}, which is defined through the Wigner transform of the density matrix as
\begin{equation}
W_{nn'}(\boldsymbol{x},\boldsymbol{k},t) 
= 
\sum_{\Delta \boldsymbol{k}} e^{i 2 \Delta \boldsymbol{k} \cdot\boldsymbol{x}} \rho_{nn'} \Big(\boldsymbol{k}+\Delta \boldsymbol{k},\boldsymbol{k}- \Delta \boldsymbol{k};t \Big) \;,
\end{equation}
where we used the rotation of coordinates $\boldsymbol{k},\boldsymbol{k}' \to \frac{\boldsymbol{k}+\boldsymbol{k}'}{2}, \boldsymbol{k}'-\boldsymbol{k}$.

Note that that the position $\boldsymbol{x}$ appearing in the Wigner transform is, to  be precise, a Bravais lattice vector and thus a discrete variable.
However, when studying transport properties, we are only interested in the macroscopic behavior of the system.
Under this macroscopic limit, we only study the changes of $\boldsymbol{x}$ on a length-scale much larger than the lattice parameter, so that $\boldsymbol{x}$ can be approximated as a continuum variable.
As a result, $W$ admits a continuous derivative with respect to $\boldsymbol{x}$ and, thanks to the gauge choice on the wavefunction $W$ is also differentiable with respect to $\boldsymbol{k}$.

As demonstrated by Moyal \cite{moyal1949,Groenewold1946}, the Wigner function obeys the following equation of motion:
\begin{equation}
\frac{\partial W_{nn'}(\boldsymbol{x},\boldsymbol{k},t)}{\partial t} 
= 
- \big\{ \big\{ W(\boldsymbol{x},\boldsymbol{k},t), H(\boldsymbol{x},\boldsymbol{k}) \big\} \big\}_{nn'}
=
\frac{i}{\hbar} \big( W(\boldsymbol{x},\boldsymbol{k},t) \star H(\boldsymbol{x},\boldsymbol{k}) - H(\boldsymbol{x},\boldsymbol{k}) \star W(\boldsymbol{x},\boldsymbol{k},t) \big)_{nn'} 
\;,
\end{equation}
where $\{\{f,g\}\}$ indicates the Moyal bracket between two operators $f$ and $g$, and we defined the Moyal product $\star$ as:
\begin{equation}
f \star g 
= 
f(\boldsymbol{x},\boldsymbol{k}) 
\exp\bigg(\frac{i}{2} \bigg( \frac{\cev{\partial}}{\partial \boldsymbol{x}} \cdot \frac{\vec{\partial}}{\partial \boldsymbol{k}} - \frac{\cev{\partial}}{\partial \boldsymbol{k}} \cdot \frac{\vec{\partial}}{\partial \boldsymbol{x}} \bigg) \bigg) 
g(\boldsymbol{x},\boldsymbol{k}) \;,
\end{equation}
where the left/right arrow indicates that the derivative operator acts on the operator to the left/right.

The equation of motion for $W$ is the phase-space analogous of the Liouville-Von Neumann equation of motion for the density matrix, and therefore has a complexity comparable to that of  Schroedinger's equation.

Now, we can further simplify this equation of motion by making the hypothesis that both $H$ and $W$ are slowly varying functions of $\boldsymbol{x}$ and $\boldsymbol{k}$.
We then expand the exponential appearing in the Moyal product in Taylor series and approximate the equation of motion as
\begin{align}
\frac{\partial W_{nn'}(\boldsymbol{x},\boldsymbol{k},t)}{\partial t} 
%= &
%\frac{i}{\hbar} \bigg( 
%W(\boldsymbol{x},\boldsymbol{k},t) \exp\bigg(\frac{i\hbar}{2} \bigg( \frac{\cev{\partial}}{\partial \boldsymbol{x}} \frac{\vec{\partial}}{\partial \boldsymbol{k}} - \frac{\cev{\partial}}{\partial \boldsymbol{k}} \frac{\vec{\partial}}{\partial \boldsymbol{x}} \bigg) \bigg) 
% H(\boldsymbol{x},\boldsymbol{k}) - H(\boldsymbol{x},\boldsymbol{k}) \exp\bigg(\frac{i\hbar}{2} \bigg( \frac{\cev{\partial}}{\partial \boldsymbol{x}} \frac{\vec{\partial}}{\partial \boldsymbol{k}} - \frac{\cev{\partial}}{\partial \boldsymbol{k}} \frac{\vec{\partial}}{\partial \boldsymbol{x}} \bigg) \bigg) 
% W(\boldsymbol{x},\boldsymbol{k},t) \bigg) 
%\\
%
%
%
\approx &
\frac{i}{\hbar} \Big( W(\boldsymbol{x},\boldsymbol{k},t) H(\boldsymbol{x},\boldsymbol{k}) - H(\boldsymbol{x},\boldsymbol{k}) W(\boldsymbol{x},\boldsymbol{k},t) \Big)_{nn'} \nonumber \\
&- 
\frac{1}{2} \bigg(  W(\boldsymbol{x},\boldsymbol{k},t) 
\bigg( \frac{\cev{\partial}}{\partial \boldsymbol{x}} \cdot \frac{\vec{\partial}}{\partial \boldsymbol{k}} - \frac{\cev{\partial}}{\partial \boldsymbol{k}} \cdot \frac{\vec{\partial}}{\partial \boldsymbol{x}} \bigg)
H(\boldsymbol{x},\boldsymbol{k})-
 H(\boldsymbol{x},\boldsymbol{k})
 \bigg( \frac{\cev{\partial}}{\partial \boldsymbol{x}} \cdot \frac{\vec{\partial}}{\partial \boldsymbol{k}} - \frac{\cev{\partial}}{\partial \boldsymbol{k}} \cdot \frac{\vec{\partial}}{\partial \boldsymbol{x}} \bigg)
  W(\boldsymbol{x},\boldsymbol{k},t)
   \bigg)_{nn'}
\\
= &
- \frac{i}{\hbar} \Big[ H(\boldsymbol{x},\boldsymbol{k}) , W(\boldsymbol{x},\boldsymbol{k},t) \Big]_{nn'}
- \frac{1}{2} \Big\{  \frac{\partial H(\boldsymbol{x},\boldsymbol{k})}{\partial \boldsymbol{k}} , \cdot \frac{\partial W(\boldsymbol{x},\boldsymbol{k},t)}{\partial \boldsymbol{x}} \Big \}_{nn'}
+ \frac{1}{2} \Big\{ \frac{\partial H(\boldsymbol{x},\boldsymbol{k})}{\partial \boldsymbol{x}} , \cdot \frac{\partial W(\boldsymbol{x},\boldsymbol{k},t)}{\partial \boldsymbol{k}} \Big \}_{nn'} \;.
\end{align}
Note that, if $H$ and $W$ commute (for example, if the two are diagonal in the band index $n$), this equation reduces to the Poisson bracket, i.e. the time evolution of a classical Hamiltonian.
 
The equation of motion is almost in the final form reported in the main text.
However, it is still expressed in terms of the basis set $\ket{\tilde{\psi}_{\boldsymbol{k}n}}$.
While convenient for the derivation, it is more practical to work with an equation in terms of the Bloch index $b$, rather than the Wannier index $n$ (the Wannier function basis set doesn't in general diagonalize the Bloch Hamiltonian).
Therefore, we rotate results in the $\ket{\psi_{\boldsymbol{k}bs}}$ basis set and write $W$ and $H$ as:
\begin{equation}
H_{bb'ss'}(\boldsymbol{x},\boldsymbol{k}) = \sum_{nn'} U^{\dagger}_{bs n}(\boldsymbol{k}) H_{nn'}(\boldsymbol{x},\boldsymbol{k})  U_{b's'n'}(\boldsymbol{k})
\;,
\end{equation}
and 
\begin{equation}
f_{bb'ss'}(\boldsymbol{x},\boldsymbol{k}) = \sum_{nn'} U^{\dagger}_{bs n}(\boldsymbol{k}) W_{nn'}(\boldsymbol{x},\boldsymbol{k})  U_{b's'n'}(\boldsymbol{k})
\;.
\end{equation}
The equation of motion can thus be written as:
\begin{equation}
\frac{\partial f_{bb'ss'}(\boldsymbol{x},\boldsymbol{k},t)}{\partial t} 
=
- i \Big[ H(\boldsymbol{x},\boldsymbol{k}) , f(\boldsymbol{x},\boldsymbol{k},t) \Big]_{bb'ss'}
- \frac{1}{2} \Big\{  \frac{\partial H(\boldsymbol{x},\boldsymbol{k})}{\partial \boldsymbol{k}} , \cdot \frac{\partial f(\boldsymbol{x},\boldsymbol{k},t)}{\partial \boldsymbol{x}} \Big \}_{bb'ss'}
+ \frac{1}{2} \Big\{ \frac{\partial H(\boldsymbol{x},\boldsymbol{k})}{\partial \boldsymbol{x}} , \cdot \frac{\partial f(\boldsymbol{x},\boldsymbol{k},t)}{\partial \boldsymbol{k}} \Big \}_{bb'ss'} \;.
\end{equation}

We now want to manipulate the matrix elements of the Hamiltonian entering the equation of motion for the Wigner function.
First, we note that the Wigner transform of the Hamiltonian at Eq. \ref{hamiltonian_supp} is Eq. \ref{hamiltonian_supp} itself, because $H_0$ (a Bloch Hamiltonian) is diagonal in $\boldsymbol{k}$ and the coupling with the electric-field is diagonal in $\boldsymbol{x}$. The matrix elements of such Hamiltonian are
\begin{equation}
\Braket{\psi_{\boldsymbol{k}bs} | H(\boldsymbol{x},\boldsymbol{k}) | \psi_{\boldsymbol{k}b's'}}
=
\epsilon_{bs}(\boldsymbol{k}) \delta_{bb'} \delta_{ss'}
+
\boldsymbol{d}_{bb'ss'}(\boldsymbol{k}) \cdot \boldsymbol{E}
=
[\mathcal{E}(\boldsymbol{k}) + \boldsymbol{D}(\boldsymbol{k})\cdot\boldsymbol{E}]_{bb'ss'} \;,
\end{equation}
where we introduced two tensors $\mathcal{E}(\boldsymbol{k})$ and $\boldsymbol{D}(\boldsymbol{k})$ containing the single-particle energies $\epsilon_{bs}(\boldsymbol{k})$ and dipoles $\boldsymbol{d}_{bb'ss'}(\boldsymbol{k})$. 
The dipole operator requires some care, since the position operator is not well-defined in a periodic system.
The off-diagonal terms $b \neq b'$ and $s\neq s'$ satisfy:
\begin{equation}
\boldsymbol{d}_{bb'ss'}(\boldsymbol{k})
=
\Braket{ \psi_{\boldsymbol{k}bs} | e \boldsymbol{x} | \psi_{\boldsymbol{k}b's'}} 
= 
e \frac{\Braket{ \psi_{\boldsymbol{k}bs} | [H^0,\boldsymbol{x}] | \psi_{\boldsymbol{k}b's'}}}{\epsilon_{bs}(\boldsymbol{k})-\epsilon_{b's'}(\boldsymbol{k})} 
= 
- i e \frac{\boldsymbol{v}_{bb'ss'}(\boldsymbol{k})}{\epsilon_{bs}(\boldsymbol{k})-\epsilon_{b's'}(\boldsymbol{k})}  , \quad \text{for }b \neq b' \text{ and } s\neq s'\;,
\end{equation}
where $\boldsymbol{v}_{bb'ss'}(\boldsymbol{k})$ is the velocity operator.
The diagonal terms $d_{bbss}(\boldsymbol{k})$ are ill-defined \cite{BLOUNT1962305}.
Luckily, these terms appear only inside a commutator, so that the diagonal terms don't contribute.
We thus set $d_{bbss}(\boldsymbol{k})=0$ without altering results.

The derivatives of the Hamiltonian are readily computed as
\begin{equation}
\Braket{\psi_{\boldsymbol{k}bs} | \frac{\partial H(\boldsymbol{x},\boldsymbol{k}) }{\partial \boldsymbol{x}} | \psi_{\boldsymbol{k}b's'}}
=
e \boldsymbol{E} \delta_{bb'} \delta_{ss'}\;,
\end{equation}
and
\begin{equation}
\Braket{\psi_{\boldsymbol{k}bs} 
|
\frac{\partial H(\boldsymbol{x},\boldsymbol{k}) }{\partial \boldsymbol{k}}
|
\psi_{\boldsymbol{k}b's'}}
=
v_{bb'ss'}(\boldsymbol{k}) \;.
\end{equation}

Combining all this terms together, the equation of motion for the Wigner function $f_{bb'ss'}(\boldsymbol{x},\boldsymbol{k},t)$ is
\begin{equation}
\frac{\partial f_{bb'ss'}(\boldsymbol{x},\boldsymbol{k},t)}{\partial t}
+
 \frac{i}{\hbar} \Big[ \mathcal{E}(\boldsymbol{k}) + \boldsymbol{D}(\boldsymbol{k})\cdot\boldsymbol{E} , f(\boldsymbol{x},\boldsymbol{k},t) \Big]_{bb'ss'}
+
 \frac{1}{2} \Big\{ \boldsymbol{v}(\boldsymbol{k}) , \cdot \frac{\partial f(\boldsymbol{x},\boldsymbol{k},t)}{\partial \boldsymbol{x}} \Big \}_{bb'ss'}
-
 e \boldsymbol{E} \cdot \frac{\partial f_{bb'ss'}(\boldsymbol{x},\boldsymbol{k},t)}{\partial \boldsymbol{k}}
=
0
\;.
\end{equation}

This is the equation of motion for the Hamiltonian $H_0 + e \boldsymbol{x}\cdot \boldsymbol{E}$, which, however, doesn't take into account for the effect of electronic collisions, in particular electron-phonon scattering.
This effect is added as a perturbation, and we define the electron-phonon collision matrix as \cite{simoncelli2019, PhysRevB.96.115420, Zhan2016, NEDJALKOV2013, Nedjalkov2011}:
\begin{equation}
\frac{\partial f_{bb'ss'}(\boldsymbol{x},\boldsymbol{k},t)}{\partial t} \bigg|_{coll} 
=
(1-\delta_{bb'}) (1-\delta_{ss'}) \frac{\Gamma_{bs}(\boldsymbol{k}) + \Gamma_{b's'}(\boldsymbol{k})}{2} f_{bb'ss'}(\boldsymbol{x},\boldsymbol{k},t)
+
\delta_{bb'} \delta_{ss'} \frac{1}{V}
\sum_{\boldsymbol{k}''b''s''} A_{\boldsymbol{k}bs,\boldsymbol{k}''b''s''} f_{b''b''s''s''}(\boldsymbol{x},\boldsymbol{k}'',t) \;.
\end{equation}
Here, the diagonal terms of $f$ are modified by the scattering matrix $A$, which is built as the electron-phonon collision matrix of the Boltzmann transport equation.
The off-diagonal terms instead are built \cite{PhysRevB.70.125324, simoncelli2019, RevModPhys.62.745} from the electron-phonon linewidths $\Gamma_{bs}(\boldsymbol{k}) = A_{\boldsymbol{k}bs,\boldsymbol{k}bs}$.
The electron-phonon scattering matrix is computed as \cite{ziman1960electrons}:
\begin{align}
A_{\boldsymbol{k}bs,\boldsymbol{k}'b's'}
=&
\delta_{\boldsymbol{k}\boldsymbol{k}'}
\delta_{bb'}
\delta_{ss'}
\frac{2\pi}{N_{\boldsymbol{q}}} \sum_{b'' s'' \nu \boldsymbol{q}}
|g_{b''bs''s\nu}(\boldsymbol{k},\boldsymbol{q})|^2
\Big[ 
\big( 1 - \bar{f}_{b''s''}(\boldsymbol{k+q}) + \bar{n}_{\nu}(\boldsymbol{q}) \big) \delta \big(\epsilon_{bs}(\boldsymbol{k})- \epsilon_{b''s''}(\boldsymbol{k+q}) - \omega_{\nu}(\boldsymbol{q})  \big) \nonumber \\
& 
\quad \quad \quad  \quad \quad \quad \quad
+ \big(\bar{f}_{b''s''}(\boldsymbol{k+q}) + \bar{n}_{\nu}(\boldsymbol{q}) \big) \delta \big( \epsilon_{bs}(\boldsymbol{k}) - \epsilon_{b''s''}(\boldsymbol{k+q}) + \omega_{\nu}(\boldsymbol{q})  \big)
\Big]
\nonumber \\
&+
\frac{2\pi}{ N_{\boldsymbol{q}}} \sum_{\nu \boldsymbol{q}}
|g_{b'bs's\nu}(\boldsymbol{k},\boldsymbol{q})|^2
\Big[ 
\bar{f}_{bs}(\boldsymbol{k}) \big(1 - \bar{f}_{b's'}(\boldsymbol{k+q}) \big) \bar{n}_{\nu}(\boldsymbol{q}) 
\delta \big( \epsilon_{bs}(\boldsymbol{k}) - \epsilon_{b's'}(\boldsymbol{k+q}) +  \omega_{\nu}(\boldsymbol{q})  \big)   \nonumber \\
&
\quad \quad \quad \quad \quad
+ 
\bar{f}_{b's'}(\boldsymbol{k+q}) \big( 1 - \bar{f}_{bs}(\boldsymbol{k}) \big) \bar{n}_{\nu}(\boldsymbol{q}) 
\delta \big( \epsilon_{bs}(\boldsymbol{k}) - \epsilon_{b's'}(\boldsymbol{k+q}) -  \omega_{\nu}(\boldsymbol{q})  \big)
\Big] \;,
\end{align}
where $\omega_{\nu}(\boldsymbol{q})$ is the phonon frequency at wavevector $\boldsymbol{q}$ and branch index $\nu$, $\bar{n}_{\nu}(\boldsymbol{q})$ is the Bose--Einstein distribution function, and $|g_{b'bs's\nu}(\boldsymbol{k},\boldsymbol{q})|^2$ is the strength of the electron-phonon interaction.
The long-range Froehlich interaction is taken into account following the procedure described in Ref. \cite{PhysRevLett.115.176401}.
All these quantities can be computed using density-functional perturbation theory.
Note that the linewidth $\Gamma_{bs}(\boldsymbol{k})$ of a carrier is readily obtained from the diagonal components of the scattering matrix $A$:
\begin{align}
\Gamma_{bs}(\boldsymbol{k}) = A_{\boldsymbol{k}bs,\boldsymbol{k}bs}
=&
\frac{2\pi}{N_{\boldsymbol{q}}} \sum_{b' s' \nu \boldsymbol{q}}
|g_{b'bs's\nu}(\boldsymbol{k},\boldsymbol{q})|^2
\Big[ 
\big( 1 - \bar{f}_{b's'}(\boldsymbol{k+q}) + \bar{n}_{\nu}(\boldsymbol{q}) \big) \delta \big(\epsilon_{bs}(\boldsymbol{k})- \epsilon_{b's'}(\boldsymbol{k+q}) - \omega_{\nu}(\boldsymbol{q})  \big) \nonumber \\
& 
\quad \quad \quad  \quad \quad \quad \quad
+ \big(\bar{f}_{b's'}(\boldsymbol{k+q}) + \bar{n}_{\nu}(\boldsymbol{q}) \big) \delta \big( \epsilon_{bs}(\boldsymbol{k}) - \epsilon_{b's'}(\boldsymbol{k+q}) + \omega_{\nu}(\boldsymbol{q})  \big)
\Big]
\end{align}
It's also worth noting that the carrier's lifetime is simply the inverse of its linewidth, as defined by the relation $\Gamma_{bs}(\boldsymbol{k})\tau_{bs}(\boldsymbol{k})=\hbar$.

Correcting the equation with the scattering operator, we finally find the aiWTE:
\begin{align}
\begin{split}
\frac{\partial f_{bb'ss'}(\boldsymbol{x},\boldsymbol{k},t)}{\partial t}
&+
 \frac{i}{\hbar} \Big[ \mathcal{E}(\boldsymbol{k}) + \boldsymbol{D}(\boldsymbol{k})\cdot\boldsymbol{E} , f(\boldsymbol{x},\boldsymbol{k},t) \Big]_{bb'ss'}
+
 \frac{1}{2} \Big\{ \boldsymbol{v}(\boldsymbol{k}) , \cdot \frac{\partial f(\boldsymbol{x},\boldsymbol{k},t)}{\partial \boldsymbol{x}} \Big \}_{bb'ss'} - \\
&-
 e \boldsymbol{E} \cdot \frac{\partial f_{bb'ss'}(\boldsymbol{x},\boldsymbol{k},t)}{\partial \boldsymbol{k}}
=
-\frac{\partial f_{bb'ss'}(\boldsymbol{x},\boldsymbol{k},t)}{\partial t} \bigg|_{coll} 
\;.
\end{split}
\end{align}

%%%%%%%%%%%%%%%%%%%%%%%%%%%%%%%%%%%%%%%%%%%%%%%%%%%%%%%%%%%%%%%%%%%%%%%%%%%%%%%%%%%%%

\subsection{Supplementary transport properties of Bi$_2$Se$_3$}

\begin{figure}
\includegraphics[scale=0.5]{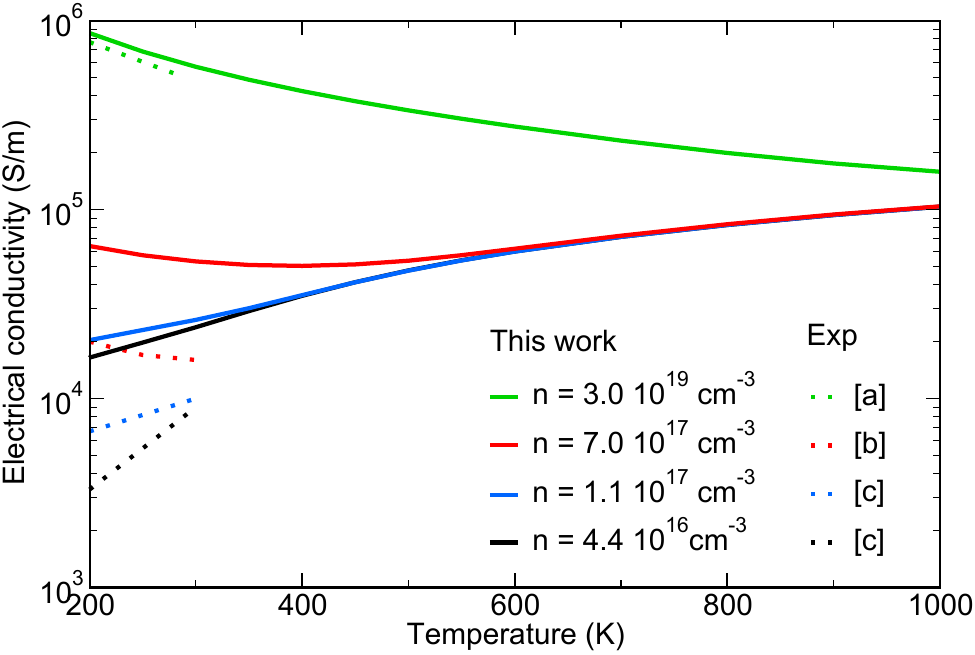}
\caption{Computational estimates of the in-plane electrical conductivity of Bi$_2$Se$_3$ as a function of temperature, for different values of electron doping concentration using the Wigner transport equation (solid lines).
We contrast these results against the experimental in-plane conductivity of single-crystals at negative doping concentrations of $3\times 10^{19}$cm$^{-3}$ \cite{doi:10.1063/1.4727957} (a), $7\times 10^{17}$cm$^{-3}$ \cite{PhysRevB.84.075316} (b), $4.4\times 10^{16}$cm$^{-3}$ and $1.1\times 10^{17}$cm$^{-3}$ \cite{natphys.bi2se3} (c).
}
\label{fig2_exp}
\end{figure}

In Fig. \ref{fig2_exp} we compare our model to available experimental data for bulk electrical conductivity.
At high doping concentrations, our results accurately reproduce the experimental measurement. 
At lower doping concentrations, the discrepancy between our simulations and experimental results increases, although qualitative trends appear still reproduced.
It must be noted that there are several factors that might introduce discrepancies between experimental and simulation results.
To mention some, the dependence of doping concentration on temperature is unclear and requires effort both in terms of modeling, as well as detailed temperature-dependent measurements of the Hall effect. 
Our simulations ignore electron-defect scattering, which may play a role especially at low temperatures, and the scattering operator is evaluated within the relaxation time approximation, which may underestimate the electrical conductivity.
Both experimental and theoretical values are affected by errors which may be hard to quantify (e.g. the dependence of computational results on the exchange-correlation functional chosen for the density functional theory calculation).
As another source of errors, recent works (e.g. \cite{PhysRevLett.117.226801, CAO2020100172}) have pointed out that the temperature dependence of the band structure may substantially change the precise values of transport coefficients, an effect that for simplicity has not been included in this work.
Nevertheless, we emphasize that the qualitative trends discussed in the main text related to the off-diagonal corrections are robust and are not significantly altered by these considerations.

%\bibliographystyle{apsrev4-1}
%\bibliography{biblio}

\begin{thebibliography}{45}%
\makeatletter
\providecommand \@ifxundefined [1]{%
 \@ifx{#1\undefined}
}%
\providecommand \@ifnum [1]{%
 \ifnum #1\expandafter \@firstoftwo
 \else \expandafter \@secondoftwo
 \fi
}%
\providecommand \@ifx [1]{%
 \ifx #1\expandafter \@firstoftwo
 \else \expandafter \@secondoftwo
 \fi
}%
\providecommand \natexlab [1]{#1}%
\providecommand \enquote  [1]{``#1''}%
\providecommand \bibnamefont  [1]{#1}%
\providecommand \bibfnamefont [1]{#1}%
\providecommand \citenamefont [1]{#1}%
\providecommand \href@noop [0]{\@secondoftwo}%
\providecommand \href [0]{\begingroup \@sanitize@url \@href}%
\providecommand \@href[1]{\@@startlink{#1}\@@href}%
\providecommand \@@href[1]{\endgroup#1\@@endlink}%
\providecommand \@sanitize@url [0]{\catcode `\\12\catcode `\$12\catcode
  `\&12\catcode `\#12\catcode `\^12\catcode `\_12\catcode `\%12\relax}%
\providecommand \@@startlink[1]{}%
\providecommand \@@endlink[0]{}%
\providecommand \url  [0]{\begingroup\@sanitize@url \@url }%
\providecommand \@url [1]{\endgroup\@href {#1}{\urlprefix }}%
\providecommand \urlprefix  [0]{URL }%
\providecommand \Eprint [0]{\href }%
\providecommand \doibase [0]{http://dx.doi.org/}%
\providecommand \selectlanguage [0]{\@gobble}%
\providecommand \bibinfo  [0]{\@secondoftwo}%
\providecommand \bibfield  [0]{\@secondoftwo}%
\providecommand \translation [1]{[#1]}%
\providecommand \BibitemOpen [0]{}%
\providecommand \bibitemStop [0]{}%
\providecommand \bibitemNoStop [0]{.\EOS\space}%
\providecommand \EOS [0]{\spacefactor3000\relax}%
\providecommand \BibitemShut  [1]{\csname bibitem#1\endcsname}%
\let\auto@bib@innerbib\@empty
%</preamble>
\bibitem [{\citenamefont {Vandecasteele}\ \emph {et~al.}(2010)\citenamefont
  {Vandecasteele}, \citenamefont {Barreiro}, \citenamefont {Lazzeri},
  \citenamefont {Bachtold},\ and\ \citenamefont {Mauri}}]{PhysRevB.82.045416}%
  \BibitemOpen
  \bibfield  {author} {\bibinfo {author} {\bibfnamefont {Niels}\ \bibnamefont
  {Vandecasteele}}, \bibinfo {author} {\bibfnamefont {Amelia}\ \bibnamefont
  {Barreiro}}, \bibinfo {author} {\bibfnamefont {Michele}\ \bibnamefont
  {Lazzeri}}, \bibinfo {author} {\bibfnamefont {Adrian}\ \bibnamefont
  {Bachtold}}, \ and\ \bibinfo {author} {\bibfnamefont {Francesco}\
  \bibnamefont {Mauri}},\ }\bibfield  {title} {\enquote {\bibinfo {title}
  {Current-voltage characteristics of graphene devices: Interplay between
  zener-klein tunneling and defects},}\ }\href {\doibase
  10.1103/PhysRevB.82.045416} {\bibfield  {journal} {\bibinfo  {journal} {Phys.
  Rev. B}\ }\textbf {\bibinfo {volume} {82}},\ \bibinfo {pages} {045416}
  (\bibinfo {year} {2010})}\BibitemShut {NoStop}%
\bibitem [{\citenamefont {Zener}(1934)}]{zener_1934}%
  \BibitemOpen
  \bibfield  {author} {\bibinfo {author} {\bibfnamefont {Clarence}\
  \bibnamefont {Zener}},\ }\bibfield  {title} {\enquote {\bibinfo {title} {A
  theory of the electrical breakdown of solid dielectrics},}\ }\href {\doibase
  10.1098/rspa.1934.0116} {\bibfield  {journal} {\bibinfo  {journal} {Proc. R.
  Soc. Lond. A}\ }\textbf {\bibinfo {volume} {145}},\ \bibinfo {pages} {523}
  (\bibinfo {year} {1934})}\BibitemShut {NoStop}%
\bibitem [{\citenamefont {Yan}\ and\ \citenamefont {Zhang}(2012)}]{Yan_2012}%
  \BibitemOpen
  \bibfield  {author} {\bibinfo {author} {\bibfnamefont {Binghai}\ \bibnamefont
  {Yan}}\ and\ \bibinfo {author} {\bibfnamefont {Shou-Cheng}\ \bibnamefont
  {Zhang}},\ }\bibfield  {title} {\enquote {\bibinfo {title} {Topological
  materials},}\ }\href {\doibase 10.1088/0034-4885/75/9/096501} {\bibfield
  {journal} {\bibinfo  {journal} {Reports on Progress in Physics}\ }\textbf
  {\bibinfo {volume} {75}},\ \bibinfo {pages} {096501} (\bibinfo {year}
  {2012})}\BibitemShut {NoStop}%
\bibitem [{\citenamefont {Armitage}\ \emph {et~al.}(2018)\citenamefont
  {Armitage}, \citenamefont {Mele},\ and\ \citenamefont
  {Vishwanath}}]{RevModPhys.90.015001}%
  \BibitemOpen
  \bibfield  {author} {\bibinfo {author} {\bibfnamefont {N.~P.}\ \bibnamefont
  {Armitage}}, \bibinfo {author} {\bibfnamefont {E.~J.}\ \bibnamefont {Mele}},
  \ and\ \bibinfo {author} {\bibfnamefont {Ashvin}\ \bibnamefont
  {Vishwanath}},\ }\bibfield  {title} {\enquote {\bibinfo {title} {Weyl and
  dirac semimetals in three-dimensional solids},}\ }\href {\doibase
  10.1103/RevModPhys.90.015001} {\bibfield  {journal} {\bibinfo  {journal}
  {Rev. Mod. Phys.}\ }\textbf {\bibinfo {volume} {90}},\ \bibinfo {pages}
  {015001} (\bibinfo {year} {2018})}\BibitemShut {NoStop}%
\bibitem [{\citenamefont {M\"uchler}\ \emph {et~al.}(2013)\citenamefont
  {M\"uchler}, \citenamefont {Casper}, \citenamefont {Yan}, \citenamefont
  {Chadov},\ and\ \citenamefont {Felser}}]{doi:10.1002/pssr.201206411}%
  \BibitemOpen
  \bibfield  {author} {\bibinfo {author} {\bibfnamefont {Lukas}\ \bibnamefont
  {M\"uchler}}, \bibinfo {author} {\bibfnamefont {Frederick}\ \bibnamefont
  {Casper}}, \bibinfo {author} {\bibfnamefont {Binghai}\ \bibnamefont {Yan}},
  \bibinfo {author} {\bibfnamefont {Stanislav}\ \bibnamefont {Chadov}}, \ and\
  \bibinfo {author} {\bibfnamefont {Claudia}\ \bibnamefont {Felser}},\
  }\bibfield  {title} {\enquote {\bibinfo {title} {Topological insulators and
  thermoelectric materials},}\ }\href {\doibase 10.1002/pssr.201206411}
  {\bibfield  {journal} {\bibinfo  {journal} {Phys. Status Solidi RRL}\
  }\textbf {\bibinfo {volume} {7}},\ \bibinfo {pages} {91--100} (\bibinfo
  {year} {2013})}\BibitemShut {NoStop}%
\bibitem [{\citenamefont {Heremans}\ \emph {et~al.}(2017)\citenamefont
  {Heremans}, \citenamefont {Cava},\ and\ \citenamefont
  {Samarth}}]{Heremans2017}%
  \BibitemOpen
  \bibfield  {author} {\bibinfo {author} {\bibfnamefont {Joseph~P.}\
  \bibnamefont {Heremans}}, \bibinfo {author} {\bibfnamefont {Robert~J.}\
  \bibnamefont {Cava}}, \ and\ \bibinfo {author} {\bibfnamefont {Nitin}\
  \bibnamefont {Samarth}},\ }\bibfield  {title} {\enquote {\bibinfo {title}
  {Tetradymites as thermoelectrics and topological insulators},}\ }\href
  {\doibase 10.1038/natrevmats.2017.49} {\bibfield  {journal} {\bibinfo
  {journal} {Nat. Rev. Mater.}\ }\textbf {\bibinfo {volume} {2}},\ \bibinfo
  {pages} {17049} (\bibinfo {year} {2017})}\BibitemShut {NoStop}%
\bibitem [{\citenamefont {Liu}\ \emph {et~al.}(2019)\citenamefont {Liu},
  \citenamefont {Williams},\ and\ \citenamefont {Cha}}]{Liu2019}%
  \BibitemOpen
  \bibfield  {author} {\bibinfo {author} {\bibfnamefont {Pengzi}\ \bibnamefont
  {Liu}}, \bibinfo {author} {\bibfnamefont {James~R.}\ \bibnamefont
  {Williams}}, \ and\ \bibinfo {author} {\bibfnamefont {Judy~J.}\ \bibnamefont
  {Cha}},\ }\bibfield  {title} {\enquote {\bibinfo {title} {Topological
  nanomaterials},}\ }\href {\doibase 10.1038/s41578-019-0113-4} {\bibfield
  {journal} {\bibinfo  {journal} {Nat. Rev. Mater.}\ }\textbf {\bibinfo
  {volume} {4}},\ \bibinfo {pages} {2058--8437} (\bibinfo {year}
  {2019})}\BibitemShut {NoStop}%
\bibitem [{\citenamefont {Wang}\ \emph {et~al.}(2011)\citenamefont {Wang},
  \citenamefont {Wang}, \citenamefont {Obukhov}, \citenamefont {Vast},
  \citenamefont {Sjakste}, \citenamefont {Tyuterev},\ and\ \citenamefont
  {Mingo}}]{PhysRevB.83.205208}%
  \BibitemOpen
  \bibfield  {author} {\bibinfo {author} {\bibfnamefont {Zhao}\ \bibnamefont
  {Wang}}, \bibinfo {author} {\bibfnamefont {Shidong}\ \bibnamefont {Wang}},
  \bibinfo {author} {\bibfnamefont {Sergey}\ \bibnamefont {Obukhov}}, \bibinfo
  {author} {\bibfnamefont {Nathalie}\ \bibnamefont {Vast}}, \bibinfo {author}
  {\bibfnamefont {Jelena}\ \bibnamefont {Sjakste}}, \bibinfo {author}
  {\bibfnamefont {Valery}\ \bibnamefont {Tyuterev}}, \ and\ \bibinfo {author}
  {\bibfnamefont {Natalio}\ \bibnamefont {Mingo}},\ }\bibfield  {title}
  {\enquote {\bibinfo {title} {Thermoelectric transport properties of silicon:
  Toward an ab initio approach},}\ }\href {\doibase 10.1103/PhysRevB.83.205208}
  {\bibfield  {journal} {\bibinfo  {journal} {Phys. Rev. B}\ }\textbf {\bibinfo
  {volume} {83}},\ \bibinfo {pages} {205208} (\bibinfo {year}
  {2011})}\BibitemShut {NoStop}%
\bibitem [{\citenamefont {Li}(2015)}]{PhysRevB.92.075405}%
  \BibitemOpen
  \bibfield  {author} {\bibinfo {author} {\bibfnamefont {Wu}~\bibnamefont
  {Li}},\ }\bibfield  {title} {\enquote {\bibinfo {title} {Electrical transport
  limited by electron-phonon coupling from boltzmann transport equation: An ab
  initio study of si, al, and ${\mathrm{mos}}_{2}$},}\ }\href {\doibase
  10.1103/PhysRevB.92.075405} {\bibfield  {journal} {\bibinfo  {journal} {Phys.
  Rev. B}\ }\textbf {\bibinfo {volume} {92}},\ \bibinfo {pages} {075405}
  (\bibinfo {year} {2015})}\BibitemShut {NoStop}%
\bibitem [{\citenamefont {Ponc\'e}\ \emph {et~al.}(2016)\citenamefont
  {Ponc\'e}, \citenamefont {Margine}, \citenamefont {Verdi},\ and\
  \citenamefont {Giustino}}]{Ponce:2016}%
  \BibitemOpen
  \bibfield  {author} {\bibinfo {author} {\bibfnamefont {S.}~\bibnamefont
  {Ponc\'e}}, \bibinfo {author} {\bibfnamefont {E.R.}\ \bibnamefont {Margine}},
  \bibinfo {author} {\bibfnamefont {C.}~\bibnamefont {Verdi}}, \ and\ \bibinfo
  {author} {\bibfnamefont {F.}~\bibnamefont {Giustino}},\ }\bibfield  {title}
  {\enquote {\bibinfo {title} {Epw: Electron-phonon coupling, transport and
  superconducting properties using maximally localized wannier functions},}\
  }\href {\doibase https://doi.org/10.1016/j.cpc.2016.07.028} {\bibfield
  {journal} {\bibinfo  {journal} {Comput. Phys. Comm.}\ }\textbf {\bibinfo
  {volume} {209}},\ \bibinfo {pages} {116--133} (\bibinfo {year}
  {2016})}\BibitemShut {NoStop}%
\bibitem [{\citenamefont {Ponc\'e}\ \emph {et~al.}(2018)\citenamefont
  {Ponc\'e}, \citenamefont {Margine},\ and\ \citenamefont
  {Giustino}}]{PhysRevB.97.121201}%
  \BibitemOpen
  \bibfield  {author} {\bibinfo {author} {\bibfnamefont {Samuel}\ \bibnamefont
  {Ponc\'e}}, \bibinfo {author} {\bibfnamefont {Elena~R.}\ \bibnamefont
  {Margine}}, \ and\ \bibinfo {author} {\bibfnamefont {Feliciano}\ \bibnamefont
  {Giustino}},\ }\bibfield  {title} {\enquote {\bibinfo {title} {Towards
  predictive many-body calculations of phonon-limited carrier mobilities in
  semiconductors},}\ }\href {\doibase 10.1103/PhysRevB.97.121201} {\bibfield
  {journal} {\bibinfo  {journal} {Phys. Rev. B}\ }\textbf {\bibinfo {volume}
  {97}},\ \bibinfo {pages} {121201} (\bibinfo {year} {2018})}\BibitemShut
  {NoStop}%
\bibitem [{\citenamefont {Fiorentini}\ and\ \citenamefont
  {Bonini}(2016)}]{PhysRevB.94.085204}%
  \BibitemOpen
  \bibfield  {author} {\bibinfo {author} {\bibfnamefont {Mattia}\ \bibnamefont
  {Fiorentini}}\ and\ \bibinfo {author} {\bibfnamefont {Nicola}\ \bibnamefont
  {Bonini}},\ }\bibfield  {title} {\enquote {\bibinfo {title} {Thermoelectric
  coefficients of $n$-doped silicon from first principles via the solution of
  the boltzmann transport equation},}\ }\href {\doibase
  10.1103/PhysRevB.94.085204} {\bibfield  {journal} {\bibinfo  {journal} {Phys.
  Rev. B}\ }\textbf {\bibinfo {volume} {94}},\ \bibinfo {pages} {085204}
  (\bibinfo {year} {2016})}\BibitemShut {NoStop}%
\bibitem [{\citenamefont {Samsonidze}\ and\ \citenamefont
  {Kozinsky}(2018)}]{doi:10.1002/aenm.201870095}%
  \BibitemOpen
  \bibfield  {author} {\bibinfo {author} {\bibfnamefont {Georgy}\ \bibnamefont
  {Samsonidze}}\ and\ \bibinfo {author} {\bibfnamefont {Boris}\ \bibnamefont
  {Kozinsky}},\ }\bibfield  {title} {\enquote {\bibinfo {title} {Thermoelectric
  materials: Accelerated screening of thermoelectric materials by
  first-principles computations of electron-phonon scattering},}\ }\href
  {\doibase 10.1002/aenm.201870095} {\bibfield  {journal} {\bibinfo  {journal}
  {Advanced Energy Materials}\ }\textbf {\bibinfo {volume} {8}},\ \bibinfo
  {pages} {1870095} (\bibinfo {year} {2018})}\BibitemShut {NoStop}%
\bibitem [{\citenamefont {Nikoli\'c}\ \emph {et~al.}(2012)\citenamefont
  {Nikoli\'c}, \citenamefont {Saha}, \citenamefont {Markussen},\ and\
  \citenamefont {Thygesen}}]{reviewNEGF}%
  \BibitemOpen
  \bibfield  {author} {\bibinfo {author} {\bibfnamefont {Branislav~K.}\
  \bibnamefont {Nikoli\'c}}, \bibinfo {author} {\bibfnamefont {Kamal~K.}\
  \bibnamefont {Saha}}, \bibinfo {author} {\bibfnamefont {Troels}\ \bibnamefont
  {Markussen}}, \ and\ \bibinfo {author} {\bibfnamefont {Kristian~S.}\
  \bibnamefont {Thygesen}},\ }\bibfield  {title} {\enquote {\bibinfo {title}
  {First-principles quantum transport modeling of thermoelectricity in
  single-molecule nanojunctions with graphene nanoribbon electrodes},}\ }\href
  {\doibase 10.1007/s10825-012-0386-y} {\bibfield  {journal} {\bibinfo
  {journal} {J. Comput. Electron.}\ }\textbf {\bibinfo {volume} {11}},\
  \bibinfo {pages} {78,92} (\bibinfo {year} {2012})}\BibitemShut {NoStop}%
\bibitem [{\citenamefont {Weinbub}\ and\ \citenamefont
  {Ferry}(2018)}]{doi:10.1063/1.5046663}%
  \BibitemOpen
  \bibfield  {author} {\bibinfo {author} {\bibfnamefont {J.}~\bibnamefont
  {Weinbub}}\ and\ \bibinfo {author} {\bibfnamefont {D.~K.}\ \bibnamefont
  {Ferry}},\ }\bibfield  {title} {\enquote {\bibinfo {title} {Recent advances
  in wigner function approaches},}\ }\href {\doibase 10.1063/1.5046663}
  {\bibfield  {journal} {\bibinfo  {journal} {Applied Physics Reviews}\
  }\textbf {\bibinfo {volume} {5}},\ \bibinfo {pages} {041104} (\bibinfo {year}
  {2018})},\ \Eprint {http://arxiv.org/abs/https://doi.org/10.1063/1.5046663}
  {https://doi.org/10.1063/1.5046663} \BibitemShut {NoStop}%
\bibitem [{\citenamefont {Wigner}(1932)}]{wigner1932}%
  \BibitemOpen
  \bibfield  {author} {\bibinfo {author} {\bibfnamefont {E.}~\bibnamefont
  {Wigner}},\ }\bibfield  {title} {\enquote {\bibinfo {title} {On the quantum
  correction for thermodynamic equilibrium},}\ }\href {\doibase
  10.1103/PhysRev.40.749} {\bibfield  {journal} {\bibinfo  {journal} {Phys.
  Rev.}\ }\textbf {\bibinfo {volume} {40}},\ \bibinfo {pages} {749--759}
  (\bibinfo {year} {1932})}\BibitemShut {NoStop}%
\bibitem [{\citenamefont {Moyal}(1949)}]{moyal1949}%
  \BibitemOpen
  \bibfield  {author} {\bibinfo {author} {\bibfnamefont {J.~E.}\ \bibnamefont
  {Moyal}},\ }\bibfield  {title} {\enquote {\bibinfo {title} {Quantum mechanics
  as a statistical theory},}\ }\href {\doibase 10.1017/S0305004100000487}
  {\bibfield  {journal} {\bibinfo  {journal} {Math. Proc. Camb. Philos. Soc.}\
  }\textbf {\bibinfo {volume} {45}},\ \bibinfo {pages} {99--124} (\bibinfo
  {year} {1949})}\BibitemShut {NoStop}%
\bibitem [{\citenamefont {Groenewold}(1946)}]{Groenewold1946}%
  \BibitemOpen
  \bibfield  {author} {\bibinfo {author} {\bibfnamefont {H.J.}\ \bibnamefont
  {Groenewold}},\ }\bibfield  {title} {\enquote {\bibinfo {title} {On the
  principles of elementary quantum mechanics},}\ }\href {\doibase
  https://doi.org/10.1016/S0031-8914(46)80059-4} {\bibfield  {journal}
  {\bibinfo  {journal} {Physica}\ }\textbf {\bibinfo {volume} {12}},\ \bibinfo
  {pages} {405 -- 460} (\bibinfo {year} {1946})}\BibitemShut {NoStop}%
\bibitem [{\citenamefont {Simoncelli}\ \emph {et~al.}(2019)\citenamefont
  {Simoncelli}, \citenamefont {Marzari},\ and\ \citenamefont
  {Mauri}}]{simoncelli2019}%
  \BibitemOpen
  \bibfield  {author} {\bibinfo {author} {\bibfnamefont {Michele}\ \bibnamefont
  {Simoncelli}}, \bibinfo {author} {\bibfnamefont {Nicola}\ \bibnamefont
  {Marzari}}, \ and\ \bibinfo {author} {\bibfnamefont {Francesco}\ \bibnamefont
  {Mauri}},\ }\bibfield  {title} {\enquote {\bibinfo {title} {Unified theory of
  thermal transport in crystals and glasses},}\ }\href {\doibase
  10.1038/s41567-019-0520-x} {\bibfield  {journal} {\bibinfo  {journal} {Nature
  Phys.}\ }\textbf {\bibinfo {volume} {15}},\ \bibinfo {pages} {809--813}
  (\bibinfo {year} {2019})}\BibitemShut {NoStop}%
\bibitem [{\citenamefont {Iotti}\ \emph {et~al.}(2017)\citenamefont {Iotti},
  \citenamefont {Dolcini},\ and\ \citenamefont {Rossi}}]{PhysRevB.96.115420}%
  \BibitemOpen
  \bibfield  {author} {\bibinfo {author} {\bibfnamefont {Rita~Claudia}\
  \bibnamefont {Iotti}}, \bibinfo {author} {\bibfnamefont {Fabrizio}\
  \bibnamefont {Dolcini}}, \ and\ \bibinfo {author} {\bibfnamefont {Fausto}\
  \bibnamefont {Rossi}},\ }\bibfield  {title} {\enquote {\bibinfo {title}
  {Wigner-function formalism applied to semiconductor quantum devices: Need for
  nonlocal scattering models},}\ }\href {\doibase 10.1103/PhysRevB.96.115420}
  {\bibfield  {journal} {\bibinfo  {journal} {Phys. Rev. B}\ }\textbf {\bibinfo
  {volume} {96}},\ \bibinfo {pages} {115420} (\bibinfo {year}
  {2017})}\BibitemShut {NoStop}%
\bibitem [{\citenamefont {Zhan}\ \emph {et~al.}(2016)\citenamefont {Zhan},
  \citenamefont {Colom{\'e}s},\ and\ \citenamefont {Oriols}}]{Zhan2016}%
  \BibitemOpen
  \bibfield  {author} {\bibinfo {author} {\bibfnamefont {Z.}~\bibnamefont
  {Zhan}}, \bibinfo {author} {\bibfnamefont {E.}~\bibnamefont {Colom{\'e}s}}, \
  and\ \bibinfo {author} {\bibfnamefont {X.}~\bibnamefont {Oriols}},\
  }\bibfield  {title} {\enquote {\bibinfo {title} {Unphysical features in the
  application of the boltzmann collision operator in the time-dependent
  modeling of quantum transport},}\ }\href {\doibase 10.1007/s10825-016-0875-5}
  {\bibfield  {journal} {\bibinfo  {journal} {J. Comput. Electron.}\ }\textbf
  {\bibinfo {volume} {15}},\ \bibinfo {pages} {1206--1218} (\bibinfo {year}
  {2016})}\BibitemShut {NoStop}%
\bibitem [{\citenamefont {Nedjalkov}\ \emph {et~al.}(2013)\citenamefont
  {Nedjalkov}, \citenamefont {Selberherr}, \citenamefont {Ferry}, \citenamefont
  {Vasileska}, \citenamefont {Dollfus}, \citenamefont {Querlioz}, \citenamefont
  {Dimov},\ and\ \citenamefont {Schwaha}}]{NEDJALKOV2013}%
  \BibitemOpen
  \bibfield  {author} {\bibinfo {author} {\bibfnamefont {M.}~\bibnamefont
  {Nedjalkov}}, \bibinfo {author} {\bibfnamefont {S.}~\bibnamefont
  {Selberherr}}, \bibinfo {author} {\bibfnamefont {D.K.}\ \bibnamefont
  {Ferry}}, \bibinfo {author} {\bibfnamefont {D.}~\bibnamefont {Vasileska}},
  \bibinfo {author} {\bibfnamefont {P.}~\bibnamefont {Dollfus}}, \bibinfo
  {author} {\bibfnamefont {D.}~\bibnamefont {Querlioz}}, \bibinfo {author}
  {\bibfnamefont {I.}~\bibnamefont {Dimov}}, \ and\ \bibinfo {author}
  {\bibfnamefont {P.}~\bibnamefont {Schwaha}},\ }\bibfield  {title} {\enquote
  {\bibinfo {title} {Physical scales in the wigner-boltzmann equation},}\
  }\href {\doibase https://doi.org/10.1016/j.aop.2012.10.001} {\bibfield
  {journal} {\bibinfo  {journal} {Ann. Phys.}\ }\textbf {\bibinfo {volume}
  {328}},\ \bibinfo {pages} {220 -- 237} (\bibinfo {year} {2013})}\BibitemShut
  {NoStop}%
\bibitem [{\citenamefont {Nedjalkov}\ \emph {et~al.}(2011)\citenamefont
  {Nedjalkov}, \citenamefont {Querlioz}, \citenamefont {Dollfus},\ and\
  \citenamefont {Kosina}}]{Nedjalkov2011}%
  \BibitemOpen
  \bibfield  {author} {\bibinfo {author} {\bibfnamefont {M.}~\bibnamefont
  {Nedjalkov}}, \bibinfo {author} {\bibfnamefont {D.}~\bibnamefont {Querlioz}},
  \bibinfo {author} {\bibfnamefont {P.}~\bibnamefont {Dollfus}}, \ and\
  \bibinfo {author} {\bibfnamefont {H.}~\bibnamefont {Kosina}},\ }\enquote
  {\bibinfo {title} {Wigner function approach},}\ in\ \href {\doibase
  10.1007/978-1-4419-8840-9_5} {\emph {\bibinfo {booktitle} {Nano-Electronic
  Devices: Semiclassical and Quantum Transport Modeling}}},\ \bibinfo {editor}
  {edited by\ \bibinfo {editor} {\bibfnamefont {Dragica}\ \bibnamefont
  {Vasileska}}\ and\ \bibinfo {editor} {\bibfnamefont {Stephen~M.}\
  \bibnamefont {Goodnick}}}\ (\bibinfo  {publisher} {Springer New York},\
  \bibinfo {address} {New York, NY},\ \bibinfo {year} {2011})\ pp.\ \bibinfo
  {pages} {289--358}\BibitemShut {NoStop}%
\bibitem [{\citenamefont {Rossi}\ and\ \citenamefont
  {Kuhn}(2002)}]{RevModPhys.74.895}%
  \BibitemOpen
  \bibfield  {author} {\bibinfo {author} {\bibfnamefont {Fausto}\ \bibnamefont
  {Rossi}}\ and\ \bibinfo {author} {\bibfnamefont {Tilmann}\ \bibnamefont
  {Kuhn}},\ }\bibfield  {title} {\enquote {\bibinfo {title} {Theory of
  ultrafast phenomena in photoexcited semiconductors},}\ }\href {\doibase
  10.1103/RevModPhys.74.895} {\bibfield  {journal} {\bibinfo  {journal} {Rev.
  Mod. Phys.}\ }\textbf {\bibinfo {volume} {74}},\ \bibinfo {pages} {895--950}
  (\bibinfo {year} {2002})}\BibitemShut {NoStop}%
\bibitem [{\citenamefont {H\"ubner}\ and\ \citenamefont
  {Graham}(1996)}]{PhysRevB.53.4870}%
  \BibitemOpen
  \bibfield  {author} {\bibinfo {author} {\bibfnamefont {R.}~\bibnamefont
  {H\"ubner}}\ and\ \bibinfo {author} {\bibfnamefont {R.}~\bibnamefont
  {Graham}},\ }\bibfield  {title} {\enquote {\bibinfo {title} {Landau-zener
  transitions and dissipation in a mesoscopic ring},}\ }\href {\doibase
  10.1103/PhysRevB.53.4870} {\bibfield  {journal} {\bibinfo  {journal} {Phys.
  Rev. B}\ }\textbf {\bibinfo {volume} {53}},\ \bibinfo {pages} {4870--4885}
  (\bibinfo {year} {1996})}\BibitemShut {NoStop}%
\bibitem [{\citenamefont {Krieger}\ and\ \citenamefont
  {Iafrate}(1987)}]{PhysRevB.35.9644}%
  \BibitemOpen
  \bibfield  {author} {\bibinfo {author} {\bibfnamefont {J.~B.}\ \bibnamefont
  {Krieger}}\ and\ \bibinfo {author} {\bibfnamefont {G.~J.}\ \bibnamefont
  {Iafrate}},\ }\bibfield  {title} {\enquote {\bibinfo {title} {Quantum
  transport for bloch electrons in a spatially homogeneous electric field},}\
  }\href {\doibase 10.1103/PhysRevB.35.9644} {\bibfield  {journal} {\bibinfo
  {journal} {Phys. Rev. B}\ }\textbf {\bibinfo {volume} {35}},\ \bibinfo
  {pages} {9644--9658} (\bibinfo {year} {1987})}\BibitemShut {NoStop}%
\bibitem [{\citenamefont {Kan\'e}\ \emph {et~al.}(2012)\citenamefont {Kan\'e},
  \citenamefont {Lazzeri},\ and\ \citenamefont {Mauri}}]{PhysRevB.86.155433}%
  \BibitemOpen
  \bibfield  {author} {\bibinfo {author} {\bibfnamefont {Gaston}\ \bibnamefont
  {Kan\'e}}, \bibinfo {author} {\bibfnamefont {Michele}\ \bibnamefont
  {Lazzeri}}, \ and\ \bibinfo {author} {\bibfnamefont {Francesco}\ \bibnamefont
  {Mauri}},\ }\bibfield  {title} {\enquote {\bibinfo {title} {Zener tunneling
  in the electrical transport of quasimetallic carbon nanotubes},}\ }\href
  {\doibase 10.1103/PhysRevB.86.155433} {\bibfield  {journal} {\bibinfo
  {journal} {Phys. Rev. B}\ }\textbf {\bibinfo {volume} {86}},\ \bibinfo
  {pages} {155433} (\bibinfo {year} {2012})}\BibitemShut {NoStop}%
\bibitem [{\citenamefont {Kan\'e}\ \emph {et~al.}(2015)\citenamefont {Kan\'e},
  \citenamefont {Lazzeri},\ and\ \citenamefont {Mauri}}]{kane_2015}%
  \BibitemOpen
  \bibfield  {author} {\bibinfo {author} {\bibfnamefont {Gaston}\ \bibnamefont
  {Kan\'e}}, \bibinfo {author} {\bibfnamefont {Michele}\ \bibnamefont
  {Lazzeri}}, \ and\ \bibinfo {author} {\bibfnamefont {Francesco}\ \bibnamefont
  {Mauri}},\ }\bibfield  {title} {\enquote {\bibinfo {title} {High-field
  transport in graphene: the impact of zener tunneling},}\ }\href {\doibase
  10.1088/0953-8984/27/16/164205} {\bibfield  {journal} {\bibinfo  {journal}
  {J. Phys.: Condends. Matter}\ }\textbf {\bibinfo {volume} {27}},\ \bibinfo
  {pages} {164205} (\bibinfo {year} {2015})}\BibitemShut {NoStop}%
\bibitem [{\citenamefont {Iafrate}\ \emph {et~al.}(2017)\citenamefont
  {Iafrate}, \citenamefont {Sokolov},\ and\ \citenamefont
  {Krieger}}]{PhysRevB.96.144303}%
  \BibitemOpen
  \bibfield  {author} {\bibinfo {author} {\bibfnamefont {G.~J.}\ \bibnamefont
  {Iafrate}}, \bibinfo {author} {\bibfnamefont {V.~N.}\ \bibnamefont
  {Sokolov}}, \ and\ \bibinfo {author} {\bibfnamefont {J.~B.}\ \bibnamefont
  {Krieger}},\ }\bibfield  {title} {\enquote {\bibinfo {title} {Quantum
  transport and the wigner distribution function for bloch electrons in
  spatially homogeneous electric and magnetic fields},}\ }\href {\doibase
  10.1103/PhysRevB.96.144303} {\bibfield  {journal} {\bibinfo  {journal} {Phys.
  Rev. B}\ }\textbf {\bibinfo {volume} {96}},\ \bibinfo {pages} {144303}
  (\bibinfo {year} {2017})}\BibitemShut {NoStop}%
\bibitem [{\citenamefont {Giannozzi}\ \emph {et~al.}(2009)\citenamefont
  {Giannozzi}, \citenamefont {Baroni}, \citenamefont {Bonini}, \citenamefont
  {Calandra}, \citenamefont {Car}, \citenamefont {Cavazzoni}, \citenamefont
  {Ceresoli}, \citenamefont {Chiarotti}, \citenamefont {Cococcioni},
  \citenamefont {Dabo}, \citenamefont {Corso}, \citenamefont {de~Gironcoli},
  \citenamefont {Fabris}, \citenamefont {Fratesi}, \citenamefont {Gebauer},
  \citenamefont {Gerstmann}, \citenamefont {Gougoussis}, \citenamefont
  {Kokalj}, \citenamefont {Lazzeri}, \citenamefont {Martin-Samos},
  \citenamefont {Marzari}, \citenamefont {Mauri}, \citenamefont {Mazzarello},
  \citenamefont {Paolini}, \citenamefont {Pasquarello}, \citenamefont
  {Paulatto}, \citenamefont {Sbraccia}, \citenamefont {Scandolo}, \citenamefont
  {Sclauzero}, \citenamefont {Seitsonen}, \citenamefont {Smogunov},
  \citenamefont {Umari},\ and\ \citenamefont
  {Wentzcovitch}}]{giannozzi2009quantum}%
  \BibitemOpen
  \bibfield  {author} {\bibinfo {author} {\bibfnamefont {Paolo}\ \bibnamefont
  {Giannozzi}}, \bibinfo {author} {\bibfnamefont {Stefano}\ \bibnamefont
  {Baroni}}, \bibinfo {author} {\bibfnamefont {Nicola}\ \bibnamefont {Bonini}},
  \bibinfo {author} {\bibfnamefont {Matteo}\ \bibnamefont {Calandra}}, \bibinfo
  {author} {\bibfnamefont {Roberto}\ \bibnamefont {Car}}, \bibinfo {author}
  {\bibfnamefont {Carlo}\ \bibnamefont {Cavazzoni}}, \bibinfo {author}
  {\bibfnamefont {Davide}\ \bibnamefont {Ceresoli}}, \bibinfo {author}
  {\bibfnamefont {Guido~L}\ \bibnamefont {Chiarotti}}, \bibinfo {author}
  {\bibfnamefont {Matteo}\ \bibnamefont {Cococcioni}}, \bibinfo {author}
  {\bibfnamefont {Ismaila}\ \bibnamefont {Dabo}}, \bibinfo {author}
  {\bibfnamefont {Andrea~Dal}\ \bibnamefont {Corso}}, \bibinfo {author}
  {\bibfnamefont {Stefano}\ \bibnamefont {de~Gironcoli}}, \bibinfo {author}
  {\bibfnamefont {Stefano}\ \bibnamefont {Fabris}}, \bibinfo {author}
  {\bibfnamefont {Guido}\ \bibnamefont {Fratesi}}, \bibinfo {author}
  {\bibfnamefont {Ralph}\ \bibnamefont {Gebauer}}, \bibinfo {author}
  {\bibfnamefont {Uwe}\ \bibnamefont {Gerstmann}}, \bibinfo {author}
  {\bibfnamefont {Christos}\ \bibnamefont {Gougoussis}}, \bibinfo {author}
  {\bibfnamefont {Anton}\ \bibnamefont {Kokalj}}, \bibinfo {author}
  {\bibfnamefont {Michele}\ \bibnamefont {Lazzeri}}, \bibinfo {author}
  {\bibfnamefont {Layla}\ \bibnamefont {Martin-Samos}}, \bibinfo {author}
  {\bibfnamefont {Nicola}\ \bibnamefont {Marzari}}, \bibinfo {author}
  {\bibfnamefont {Francesco}\ \bibnamefont {Mauri}}, \bibinfo {author}
  {\bibfnamefont {Riccardo}\ \bibnamefont {Mazzarello}}, \bibinfo {author}
  {\bibfnamefont {Stefano}\ \bibnamefont {Paolini}}, \bibinfo {author}
  {\bibfnamefont {Alfredo}\ \bibnamefont {Pasquarello}}, \bibinfo {author}
  {\bibfnamefont {Lorenzo}\ \bibnamefont {Paulatto}}, \bibinfo {author}
  {\bibfnamefont {Carlo}\ \bibnamefont {Sbraccia}}, \bibinfo {author}
  {\bibfnamefont {Sandro}\ \bibnamefont {Scandolo}}, \bibinfo {author}
  {\bibfnamefont {Gabriele}\ \bibnamefont {Sclauzero}}, \bibinfo {author}
  {\bibfnamefont {Ari~P}\ \bibnamefont {Seitsonen}}, \bibinfo {author}
  {\bibfnamefont {Alexander}\ \bibnamefont {Smogunov}}, \bibinfo {author}
  {\bibfnamefont {Paolo}\ \bibnamefont {Umari}}, \ and\ \bibinfo {author}
  {\bibfnamefont {Renata~M}\ \bibnamefont {Wentzcovitch}},\ }\bibfield  {title}
  {\enquote {\bibinfo {title} {{QUANTUM ESPRESSO: a modular and open-source
  software project for quantum simulations of materials}},}\ }\href
  {https://iopscience.iop.org/article/10.1088/0953-8984/21/39/395502/pdf}
  {\bibfield  {journal} {\bibinfo  {journal} {J. Phys. Condens. Matter}\
  }\textbf {\bibinfo {volume} {21}},\ \bibinfo {pages} {395502} (\bibinfo
  {year} {2009})}\BibitemShut {NoStop}%
\bibitem [{\citenamefont {Giannozzi}\ \emph {et~al.}(2017)\citenamefont
  {Giannozzi}, \citenamefont {Andreussi}, \citenamefont {Brumme}, \citenamefont
  {Bunau}, \citenamefont {Nardelli}, \citenamefont {Calandra}, \citenamefont
  {Car}, \citenamefont {Cavazzoni}, \citenamefont {Ceresoli}, \citenamefont
  {Cococcioni}, \citenamefont {Colonna}, \citenamefont {Carnimeo},
  \citenamefont {Corso}, \citenamefont {de~Gironcoli}, \citenamefont {Delugas},
  \citenamefont {Jr}, \citenamefont {Ferretti}, \citenamefont {Floris},
  \citenamefont {Fratesi}, \citenamefont {Fugallo}, \citenamefont {Gebauer},
  \citenamefont {Gerstmann}, \citenamefont {Giustino}, \citenamefont {Gorni},
  \citenamefont {Jia}, \citenamefont {Kawamura}, \citenamefont {Ko},
  \citenamefont {Kokalj}, \citenamefont {K\"u\c{c}\"ukbenli}, \citenamefont
  {Lazzeri}, \citenamefont {Marsili}, \citenamefont {Marzari}, \citenamefont
  {Mauri}, \citenamefont {Nguyen}, \citenamefont {Nguyen}, \citenamefont {de-la
  Roza}, \citenamefont {Paulatto}, \citenamefont {Ponc\'e}, \citenamefont
  {Rocca}, \citenamefont {Sabatini}, \citenamefont {Santra}, \citenamefont
  {Schlipf}, \citenamefont {Seitsonen}, \citenamefont {Smogunov}, \citenamefont
  {Timrov}, \citenamefont {Thonhauser}, \citenamefont {Umari}, \citenamefont
  {Vast}, \citenamefont {Wu},\ and\ \citenamefont
  {Baroni}}]{giannozzi2017advanced}%
  \BibitemOpen
  \bibfield  {author} {\bibinfo {author} {\bibfnamefont {P}~\bibnamefont
  {Giannozzi}}, \bibinfo {author} {\bibfnamefont {O}~\bibnamefont {Andreussi}},
  \bibinfo {author} {\bibfnamefont {T}~\bibnamefont {Brumme}}, \bibinfo
  {author} {\bibfnamefont {O}~\bibnamefont {Bunau}}, \bibinfo {author}
  {\bibfnamefont {M~Buongiorno}\ \bibnamefont {Nardelli}}, \bibinfo {author}
  {\bibfnamefont {M}~\bibnamefont {Calandra}}, \bibinfo {author} {\bibfnamefont
  {R}~\bibnamefont {Car}}, \bibinfo {author} {\bibfnamefont {C}~\bibnamefont
  {Cavazzoni}}, \bibinfo {author} {\bibfnamefont {D}~\bibnamefont {Ceresoli}},
  \bibinfo {author} {\bibfnamefont {M}~\bibnamefont {Cococcioni}}, \bibinfo
  {author} {\bibfnamefont {N}~\bibnamefont {Colonna}}, \bibinfo {author}
  {\bibfnamefont {I}~\bibnamefont {Carnimeo}}, \bibinfo {author} {\bibfnamefont
  {A~Dal}\ \bibnamefont {Corso}}, \bibinfo {author} {\bibfnamefont
  {S}~\bibnamefont {de~Gironcoli}}, \bibinfo {author} {\bibfnamefont
  {P}~\bibnamefont {Delugas}}, \bibinfo {author} {\bibfnamefont {R~A~DiStasio}\
  \bibnamefont {Jr}}, \bibinfo {author} {\bibfnamefont {A}~\bibnamefont
  {Ferretti}}, \bibinfo {author} {\bibfnamefont {A}~\bibnamefont {Floris}},
  \bibinfo {author} {\bibfnamefont {G}~\bibnamefont {Fratesi}}, \bibinfo
  {author} {\bibfnamefont {G}~\bibnamefont {Fugallo}}, \bibinfo {author}
  {\bibfnamefont {R}~\bibnamefont {Gebauer}}, \bibinfo {author} {\bibfnamefont
  {U}~\bibnamefont {Gerstmann}}, \bibinfo {author} {\bibfnamefont
  {F}~\bibnamefont {Giustino}}, \bibinfo {author} {\bibfnamefont
  {T}~\bibnamefont {Gorni}}, \bibinfo {author} {\bibfnamefont {J}~\bibnamefont
  {Jia}}, \bibinfo {author} {\bibfnamefont {M}~\bibnamefont {Kawamura}},
  \bibinfo {author} {\bibfnamefont {H-Y}\ \bibnamefont {Ko}}, \bibinfo {author}
  {\bibfnamefont {A}~\bibnamefont {Kokalj}}, \bibinfo {author} {\bibfnamefont
  {E}~\bibnamefont {K\"u\c{c}\"ukbenli}}, \bibinfo {author} {\bibfnamefont
  {M}~\bibnamefont {Lazzeri}}, \bibinfo {author} {\bibfnamefont
  {M}~\bibnamefont {Marsili}}, \bibinfo {author} {\bibfnamefont
  {N}~\bibnamefont {Marzari}}, \bibinfo {author} {\bibfnamefont
  {F}~\bibnamefont {Mauri}}, \bibinfo {author} {\bibfnamefont {N.~L.}\
  \bibnamefont {Nguyen}}, \bibinfo {author} {\bibfnamefont {H.~V.}\
  \bibnamefont {Nguyen}}, \bibinfo {author} {\bibfnamefont {A~Otero}\
  \bibnamefont {de-la Roza}}, \bibinfo {author} {\bibfnamefont {L}~\bibnamefont
  {Paulatto}}, \bibinfo {author} {\bibfnamefont {S}~\bibnamefont {Ponc\'e}},
  \bibinfo {author} {\bibfnamefont {D}~\bibnamefont {Rocca}}, \bibinfo {author}
  {\bibfnamefont {R}~\bibnamefont {Sabatini}}, \bibinfo {author} {\bibfnamefont
  {B}~\bibnamefont {Santra}}, \bibinfo {author} {\bibfnamefont {M}~\bibnamefont
  {Schlipf}}, \bibinfo {author} {\bibfnamefont {A~P}\ \bibnamefont
  {Seitsonen}}, \bibinfo {author} {\bibfnamefont {A}~\bibnamefont {Smogunov}},
  \bibinfo {author} {\bibfnamefont {I}~\bibnamefont {Timrov}}, \bibinfo
  {author} {\bibfnamefont {T}~\bibnamefont {Thonhauser}}, \bibinfo {author}
  {\bibfnamefont {P}~\bibnamefont {Umari}}, \bibinfo {author} {\bibfnamefont
  {N}~\bibnamefont {Vast}}, \bibinfo {author} {\bibfnamefont {X}~\bibnamefont
  {Wu}}, \ and\ \bibinfo {author} {\bibfnamefont {S}~\bibnamefont {Baroni}},\
  }\bibfield  {title} {\enquote {\bibinfo {title} {{{Advanced capabilities for
  materials modelling with Quantum ESPRESSO}}},}\ }\href
  {http://stacks.iop.org/0953-8984/29/i=46/a=465901} {\bibfield  {journal}
  {\bibinfo  {journal} {J. Phys. Condens. Matter}\ }\textbf {\bibinfo {volume}
  {29}},\ \bibinfo {pages} {465901} (\bibinfo {year} {2017})}\BibitemShut
  {NoStop}%
\bibitem [{\citenamefont {Garrity}\ \emph {et~al.}(2014)\citenamefont
  {Garrity}, \citenamefont {Bennett}, \citenamefont {Rabe},\ and\ \citenamefont
  {Vanderbilt}}]{GARRITY2014446}%
  \BibitemOpen
  \bibfield  {author} {\bibinfo {author} {\bibfnamefont {Kevin~F.}\
  \bibnamefont {Garrity}}, \bibinfo {author} {\bibfnamefont {Joseph~W.}\
  \bibnamefont {Bennett}}, \bibinfo {author} {\bibfnamefont {Karin~M.}\
  \bibnamefont {Rabe}}, \ and\ \bibinfo {author} {\bibfnamefont {David}\
  \bibnamefont {Vanderbilt}},\ }\bibfield  {title} {\enquote {\bibinfo {title}
  {Pseudopotentials for high-throughput dft calculations},}\ }\href {\doibase
  https://doi.org/10.1016/j.commatsci.2013.08.053} {\bibfield  {journal}
  {\bibinfo  {journal} {Comput. Mater. Sci.}\ }\textbf {\bibinfo {volume}
  {81}},\ \bibinfo {pages} {446 -- 452} (\bibinfo {year} {2014})}\BibitemShut
  {NoStop}%
\bibitem [{\citenamefont {Schubert}\ \emph {et~al.}(1953)\citenamefont
  {Schubert}, \citenamefont {Anderko}, \citenamefont {Kluge}, \citenamefont
  {Beeskow}, \citenamefont {Ilschner}, \citenamefont {D\"orre},\ and\
  \citenamefont {Esslinger}}]{bi2se3_structure}%
  \BibitemOpen
  \bibfield  {author} {\bibinfo {author} {\bibfnamefont {K.}~\bibnamefont
  {Schubert}}, \bibinfo {author} {\bibfnamefont {K.}~\bibnamefont {Anderko}},
  \bibinfo {author} {\bibfnamefont {M.}~\bibnamefont {Kluge}}, \bibinfo
  {author} {\bibfnamefont {H.}~\bibnamefont {Beeskow}}, \bibinfo {author}
  {\bibfnamefont {M.}~\bibnamefont {Ilschner}}, \bibinfo {author}
  {\bibfnamefont {E.}~\bibnamefont {D\"orre}}, \ and\ \bibinfo {author}
  {\bibfnamefont {P.}~\bibnamefont {Esslinger}},\ }\bibfield  {title} {\enquote
  {\bibinfo {title} {Strukturuntersuchung der legierungsphasen cu$_2$te, cute,
  cu$_3$sb, inte, bi$_2$se$_3$, pd$_5$sb$_3$ und pd$_5$bi$_3$},}\ }\href
  {\doibase 10.1007/BF00590422} {\bibfield  {journal} {\bibinfo  {journal}
  {Naturwissenschaften}\ }\textbf {\bibinfo {volume} {40}},\ \bibinfo {pages}
  {269} (\bibinfo {year} {1953})}\BibitemShut {NoStop}%
\bibitem [{\citenamefont {Baroni}\ \emph {et~al.}(2001)\citenamefont {Baroni},
  \citenamefont {de~Gironcoli}, \citenamefont {Dal~Corso},\ and\ \citenamefont
  {Giannozzi}}]{baroni_revmodphys}%
  \BibitemOpen
  \bibfield  {author} {\bibinfo {author} {\bibfnamefont {Stefano}\ \bibnamefont
  {Baroni}}, \bibinfo {author} {\bibfnamefont {Stefano}\ \bibnamefont
  {de~Gironcoli}}, \bibinfo {author} {\bibfnamefont {Andrea}\ \bibnamefont
  {Dal~Corso}}, \ and\ \bibinfo {author} {\bibfnamefont {Paolo}\ \bibnamefont
  {Giannozzi}},\ }\bibfield  {title} {\enquote {\bibinfo {title} {Phonons and
  related crystal properties from density-functional perturbation theory},}\
  }\href {\doibase 10.1103/RevModPhys.73.515} {\bibfield  {journal} {\bibinfo
  {journal} {Rev. Mod. Phys.}\ }\textbf {\bibinfo {volume} {73}},\ \bibinfo
  {pages} {515--562} (\bibinfo {year} {2001})}\BibitemShut {NoStop}%
\bibitem [{\citenamefont {Pizzi}\ \emph {et~al.}(2019)\citenamefont {Pizzi},
  \citenamefont {Vitale}, \citenamefont {Arita}, \citenamefont {Bl\"ugel},\
  and\ \citenamefont {et~al.}}]{pizzi_2019}%
  \BibitemOpen
  \bibfield  {author} {\bibinfo {author} {\bibfnamefont {Giovanni}\
  \bibnamefont {Pizzi}}, \bibinfo {author} {\bibfnamefont {Valerio}\
  \bibnamefont {Vitale}}, \bibinfo {author} {\bibfnamefont {Ryotaro}\
  \bibnamefont {Arita}}, \bibinfo {author} {\bibfnamefont {Stefan}\
  \bibnamefont {Bl\"ugel}}, \ and\ \bibinfo {author} {\bibfnamefont
  {Frank~Freimuth}\ \bibnamefont {et~al.}},\ }\href@noop {} {\enquote {\bibinfo
  {title} {Wannier90 as a community code: new features and applications},}\ }
  (\bibinfo {year} {2019}),\ \Eprint {http://arxiv.org/abs/1907.09788}
  {arXiv:1907.09788 [cond-mat.mtrl-sci]} \BibitemShut {NoStop}%
\bibitem [{\citenamefont {Yates}\ \emph {et~al.}(2007)\citenamefont {Yates},
  \citenamefont {Wang}, \citenamefont {Vanderbilt},\ and\ \citenamefont
  {Souza}}]{Yates2007}%
  \BibitemOpen
  \bibfield  {author} {\bibinfo {author} {\bibfnamefont {Jonathan~R.}\
  \bibnamefont {Yates}}, \bibinfo {author} {\bibfnamefont {Xinjie}\
  \bibnamefont {Wang}}, \bibinfo {author} {\bibfnamefont {David}\ \bibnamefont
  {Vanderbilt}}, \ and\ \bibinfo {author} {\bibfnamefont {Ivo}\ \bibnamefont
  {Souza}},\ }\bibfield  {title} {\enquote {\bibinfo {title} {Spectral and
  fermi surface properties from wannier interpolation},}\ }\href {\doibase
  10.1103/PhysRevB.75.195121} {\bibfield  {journal} {\bibinfo  {journal} {Phys.
  Rev. B}\ }\textbf {\bibinfo {volume} {75}},\ \bibinfo {pages} {195121}
  (\bibinfo {year} {2007})}\BibitemShut {NoStop}%
\bibitem [{\citenamefont {Verdi}\ and\ \citenamefont
  {Giustino}(2015)}]{PhysRevLett.115.176401}%
  \BibitemOpen
  \bibfield  {author} {\bibinfo {author} {\bibfnamefont {Carla}\ \bibnamefont
  {Verdi}}\ and\ \bibinfo {author} {\bibfnamefont {Feliciano}\ \bibnamefont
  {Giustino}},\ }\bibfield  {title} {\enquote {\bibinfo {title} {Fr\"ohlich
  electron-phonon vertex from first principles},}\ }\href {\doibase
  10.1103/PhysRevLett.115.176401} {\bibfield  {journal} {\bibinfo  {journal}
  {Phys. Rev. Lett.}\ }\textbf {\bibinfo {volume} {115}},\ \bibinfo {pages}
  {176401} (\bibinfo {year} {2015})}\BibitemShut {NoStop}%
\bibitem [{\citenamefont {Togo}\ and\ \citenamefont {Tanaka}(2018)}]{spglib}%
  \BibitemOpen
  \bibfield  {author} {\bibinfo {author} {\bibfnamefont {Atsushi}\ \bibnamefont
  {Togo}}\ and\ \bibinfo {author} {\bibfnamefont {Isao}\ \bibnamefont
  {Tanaka}},\ }\href@noop {} {\enquote {\bibinfo {title} {Spglib: a software
  library for crystal symmetry search},}\ } (\bibinfo {year} {2018}),\ \Eprint
  {http://arxiv.org/abs/1808.01590} {arXiv:1808.01590 [cond-mat.mtrl-sci]}
  \BibitemShut {NoStop}%
\bibitem [{\citenamefont {Hinuma}\ \emph {et~al.}(2017)\citenamefont {Hinuma},
  \citenamefont {Pizzi}, \citenamefont {Kumagai}, \citenamefont {Oba},\ and\
  \citenamefont {Tanaka}}]{HINUMA2017140}%
  \BibitemOpen
  \bibfield  {author} {\bibinfo {author} {\bibfnamefont {Yoyo}\ \bibnamefont
  {Hinuma}}, \bibinfo {author} {\bibfnamefont {Giovanni}\ \bibnamefont
  {Pizzi}}, \bibinfo {author} {\bibfnamefont {Yu}~\bibnamefont {Kumagai}},
  \bibinfo {author} {\bibfnamefont {Fumiyasu}\ \bibnamefont {Oba}}, \ and\
  \bibinfo {author} {\bibfnamefont {Isao}\ \bibnamefont {Tanaka}},\ }\bibfield
  {title} {\enquote {\bibinfo {title} {Band structure diagram paths based on
  crystallography},}\ }\href {\doibase
  https://doi.org/10.1016/j.commatsci.2016.10.015} {\bibfield  {journal}
  {\bibinfo  {journal} {Comput. Mater. Sci.}\ }\textbf {\bibinfo {volume}
  {128}},\ \bibinfo {pages} {140 -- 184} (\bibinfo {year} {2017})}\BibitemShut
  {NoStop}%
\bibitem [{\citenamefont {Martinez}\ \emph {et~al.}(2017)\citenamefont
  {Martinez}, \citenamefont {Piot}, \citenamefont {Hakl}, \citenamefont
  {Potemski}, \citenamefont {Hor}, \citenamefont {Materna}, \citenamefont
  {Strzelecka}, \citenamefont {Hruban}, \citenamefont {Caha}, \citenamefont
  {Nov\'ak}, \citenamefont {Dubroka}, \citenamefont {Dra\v{s}ar},\ and\
  \citenamefont {Orlita}}]{bi2se3_bandgap}%
  \BibitemOpen
  \bibfield  {author} {\bibinfo {author} {\bibfnamefont {G.}~\bibnamefont
  {Martinez}}, \bibinfo {author} {\bibfnamefont {B.~A.}\ \bibnamefont {Piot}},
  \bibinfo {author} {\bibfnamefont {M.}~\bibnamefont {Hakl}}, \bibinfo {author}
  {\bibfnamefont {M.}~\bibnamefont {Potemski}}, \bibinfo {author}
  {\bibfnamefont {Y.~S.}\ \bibnamefont {Hor}}, \bibinfo {author} {\bibfnamefont
  {A.}~\bibnamefont {Materna}}, \bibinfo {author} {\bibfnamefont {S.~G.}\
  \bibnamefont {Strzelecka}}, \bibinfo {author} {\bibfnamefont
  {A.}~\bibnamefont {Hruban}}, \bibinfo {author} {\bibfnamefont
  {O.}~\bibnamefont {Caha}}, \bibinfo {author} {\bibfnamefont {J.}~\bibnamefont
  {Nov\'ak}}, \bibinfo {author} {\bibfnamefont {A.}~\bibnamefont {Dubroka}},
  \bibinfo {author} {\bibfnamefont {C.}~\bibnamefont {Dra\v{s}ar}}, \ and\
  \bibinfo {author} {\bibfnamefont {M.}~\bibnamefont {Orlita}},\ }\bibfield
  {title} {\enquote {\bibinfo {title} {Determination of the energy band gap of
  bi2se3},}\ }\href {\doibase 10.1038/s41598-017-07211-x} {\bibfield  {journal}
  {\bibinfo  {journal} {Sci. Rep.}\ }\textbf {\bibinfo {volume} {7}},\ \bibinfo
  {pages} {6891} (\bibinfo {year} {2017})}\BibitemShut {NoStop}%
\bibitem [{\citenamefont {Witting}\ \emph {et~al.}(2019)\citenamefont
  {Witting}, \citenamefont {Chasapis}, \citenamefont {Ricci}, \citenamefont
  {Peters}, \citenamefont {Heinz}, \citenamefont {Hautier},\ and\ \citenamefont
  {Snyder}}]{https://doi.org/10.1002/aelm.201800904}%
  \BibitemOpen
  \bibfield  {author} {\bibinfo {author} {\bibfnamefont {Ian~T.}\ \bibnamefont
  {Witting}}, \bibinfo {author} {\bibfnamefont {Thomas~C.}\ \bibnamefont
  {Chasapis}}, \bibinfo {author} {\bibfnamefont {Francesco}\ \bibnamefont
  {Ricci}}, \bibinfo {author} {\bibfnamefont {Matthew}\ \bibnamefont {Peters}},
  \bibinfo {author} {\bibfnamefont {Nicholas~A.}\ \bibnamefont {Heinz}},
  \bibinfo {author} {\bibfnamefont {Geoffroy}\ \bibnamefont {Hautier}}, \ and\
  \bibinfo {author} {\bibfnamefont {G.~Jeffrey}\ \bibnamefont {Snyder}},\
  }\bibfield  {title} {\enquote {\bibinfo {title} {The thermoelectric
  properties of bismuth telluride},}\ }\href {\doibase
  https://doi.org/10.1002/aelm.201800904} {\bibfield  {journal} {\bibinfo
  {journal} {Advanced Electronic Materials}\ }\textbf {\bibinfo {volume} {5}},\
  \bibinfo {pages} {1800904} (\bibinfo {year} {2019})},\ \Eprint
  {http://arxiv.org/abs/https://onlinelibrary.wiley.com/doi/pdf/10.1002/aelm.201800904}
  {https://onlinelibrary.wiley.com/doi/pdf/10.1002/aelm.201800904} \BibitemShut
  {NoStop}%
\bibitem [{\citenamefont {Wee}\ \emph {et~al.}(2010)\citenamefont {Wee},
  \citenamefont {Kozinsky}, \citenamefont {Marzari},\ and\ \citenamefont
  {Fornari}}]{PhysRevB.81.045204}%
  \BibitemOpen
  \bibfield  {author} {\bibinfo {author} {\bibfnamefont {Daehyun}\ \bibnamefont
  {Wee}}, \bibinfo {author} {\bibfnamefont {Boris}\ \bibnamefont {Kozinsky}},
  \bibinfo {author} {\bibfnamefont {Nicola}\ \bibnamefont {Marzari}}, \ and\
  \bibinfo {author} {\bibfnamefont {Marco}\ \bibnamefont {Fornari}},\
  }\bibfield  {title} {\enquote {\bibinfo {title} {Effects of filling in
  ${\text{cosb}}_{3}$: Local structure, band gap, and phonons from first
  principles},}\ }\href {\doibase 10.1103/PhysRevB.81.045204} {\bibfield
  {journal} {\bibinfo  {journal} {Phys. Rev. B}\ }\textbf {\bibinfo {volume}
  {81}},\ \bibinfo {pages} {045204} (\bibinfo {year} {2010})}\BibitemShut
  {NoStop}%
\bibitem [{\citenamefont {Kim}\ \emph {et~al.}(2015)\citenamefont {Kim},
  \citenamefont {Gibbs}, \citenamefont {Tang}, \citenamefont {Wang},\ and\
  \citenamefont {Snyder}}]{doi:10.1063/1.4908244}%
  \BibitemOpen
  \bibfield  {author} {\bibinfo {author} {\bibfnamefont {Hyun-Sik}\
  \bibnamefont {Kim}}, \bibinfo {author} {\bibfnamefont {Zachary~M.}\
  \bibnamefont {Gibbs}}, \bibinfo {author} {\bibfnamefont {Yinglu}\
  \bibnamefont {Tang}}, \bibinfo {author} {\bibfnamefont {Heng}\ \bibnamefont
  {Wang}}, \ and\ \bibinfo {author} {\bibfnamefont {G.~Jeffrey}\ \bibnamefont
  {Snyder}},\ }\bibfield  {title} {\enquote {\bibinfo {title} {Characterization
  of lorenz number with seebeck coefficient measurement},}\ }\href {\doibase
  10.1063/1.4908244} {\bibfield  {journal} {\bibinfo  {journal} {APL
  Materials}\ }\textbf {\bibinfo {volume} {3}},\ \bibinfo {pages} {041506}
  (\bibinfo {year} {2015})},\ \Eprint
  {http://arxiv.org/abs/https://doi.org/10.1063/1.4908244}
  {https://doi.org/10.1063/1.4908244} \BibitemShut {NoStop}%
\bibitem [{\citenamefont {Lukas}\ \emph {et~al.}(2012)\citenamefont {Lukas},
  \citenamefont {Liu}, \citenamefont {Joshi}, \citenamefont {Zebarjadi},
  \citenamefont {Dresselhaus}, \citenamefont {Ren}, \citenamefont {Chen},\ and\
  \citenamefont {Opeil}}]{PhysRevB.85.205410}%
  \BibitemOpen
  \bibfield  {author} {\bibinfo {author} {\bibfnamefont {K.~C.}\ \bibnamefont
  {Lukas}}, \bibinfo {author} {\bibfnamefont {W.~S.}\ \bibnamefont {Liu}},
  \bibinfo {author} {\bibfnamefont {G.}~\bibnamefont {Joshi}}, \bibinfo
  {author} {\bibfnamefont {M.}~\bibnamefont {Zebarjadi}}, \bibinfo {author}
  {\bibfnamefont {M.~S.}\ \bibnamefont {Dresselhaus}}, \bibinfo {author}
  {\bibfnamefont {Z.~F.}\ \bibnamefont {Ren}}, \bibinfo {author} {\bibfnamefont
  {G.}~\bibnamefont {Chen}}, \ and\ \bibinfo {author} {\bibfnamefont {C.~P.}\
  \bibnamefont {Opeil}},\ }\bibfield  {title} {\enquote {\bibinfo {title}
  {Experimental determination of the lorenz number in
  cu$_{0.01}$bi$_{2}$te$_{2.7}$se$_{0.3}$ and bi$_{0.88}$sb$_{0.12}$},}\ }\href
  {\doibase 10.1103/PhysRevB.85.205410} {\bibfield  {journal} {\bibinfo
  {journal} {Phys. Rev. B}\ }\textbf {\bibinfo {volume} {85}},\ \bibinfo
  {pages} {205410} (\bibinfo {year} {2012})}\BibitemShut {NoStop}%
\bibitem [{\citenamefont {Chen}\ \emph {et~al.}(2019)\citenamefont {Chen},
  \citenamefont {Ma},\ and\ \citenamefont {Li}}]{PhysRevB.99.020305}%
  \BibitemOpen
  \bibfield  {author} {\bibinfo {author} {\bibfnamefont {Yani}\ \bibnamefont
  {Chen}}, \bibinfo {author} {\bibfnamefont {Jinlong}\ \bibnamefont {Ma}}, \
  and\ \bibinfo {author} {\bibfnamefont {Wu}~\bibnamefont {Li}},\ }\bibfield
  {title} {\enquote {\bibinfo {title} {Understanding the thermal conductivity
  and lorenz number in tungsten from first principles},}\ }\href {\doibase
  10.1103/PhysRevB.99.020305} {\bibfield  {journal} {\bibinfo  {journal} {Phys.
  Rev. B}\ }\textbf {\bibinfo {volume} {99}},\ \bibinfo {pages} {020305}
  (\bibinfo {year} {2019})}\BibitemShut {NoStop}%
\end{thebibliography}

\begin{thebibliography}{19}%
\makeatletter
\providecommand \@ifxundefined [1]{%
 \@ifx{#1\undefined}
}%
\providecommand \@ifnum [1]{%
 \ifnum #1\expandafter \@firstoftwo
 \else \expandafter \@secondoftwo
 \fi
}%
\providecommand \@ifx [1]{%
 \ifx #1\expandafter \@firstoftwo
 \else \expandafter \@secondoftwo
 \fi
}%
\providecommand \natexlab [1]{#1}%
\providecommand \enquote  [1]{``#1''}%
\providecommand \bibnamefont  [1]{#1}%
\providecommand \bibfnamefont [1]{#1}%
\providecommand \citenamefont [1]{#1}%
\providecommand \href@noop [0]{\@secondoftwo}%
\providecommand \href [0]{\begingroup \@sanitize@url \@href}%
\providecommand \@href[1]{\@@startlink{#1}\@@href}%
\providecommand \@@href[1]{\endgroup#1\@@endlink}%
\providecommand \@sanitize@url [0]{\catcode `\\12\catcode `\$12\catcode
  `\&12\catcode `\#12\catcode `\^12\catcode `\_12\catcode `\%12\relax}%
\providecommand \@@startlink[1]{}%
\providecommand \@@endlink[0]{}%
\providecommand \url  [0]{\begingroup\@sanitize@url \@url }%
\providecommand \@url [1]{\endgroup\@href {#1}{\urlprefix }}%
\providecommand \urlprefix  [0]{URL }%
\providecommand \Eprint [0]{\href }%
\providecommand \doibase [0]{http://dx.doi.org/}%
\providecommand \selectlanguage [0]{\@gobble}%
\providecommand \bibinfo  [0]{\@secondoftwo}%
\providecommand \bibfield  [0]{\@secondoftwo}%
\providecommand \translation [1]{[#1]}%
\providecommand \BibitemOpen [0]{}%
\providecommand \bibitemStop [0]{}%
\providecommand \bibitemNoStop [0]{.\EOS\space}%
\providecommand \EOS [0]{\spacefactor3000\relax}%
\providecommand \BibitemShut  [1]{\csname bibitem#1\endcsname}%
\let\auto@bib@innerbib\@empty
%</preamble>
\bibitem [{\citenamefont {Marzari}\ \emph {et~al.}(2012)\citenamefont
  {Marzari}, \citenamefont {Mostofi}, \citenamefont {Yates}, \citenamefont
  {Souza},\ and\ \citenamefont {Vanderbilt}}]{RevModPhys.84.1419}%
  \BibitemOpen
  \bibfield  {author} {\bibinfo {author} {\bibfnamefont {N.}~\bibnamefont
  {Marzari}}, \bibinfo {author} {\bibfnamefont {A.~A.}\ \bibnamefont
  {Mostofi}}, \bibinfo {author} {\bibfnamefont {J.~R.}\ \bibnamefont {Yates}},
  \bibinfo {author} {\bibfnamefont {I.}~\bibnamefont {Souza}}, \ and\ \bibinfo
  {author} {\bibfnamefont {D.}~\bibnamefont {Vanderbilt}},\ }\href {\doibase
  10.1103/RevModPhys.84.1419} {\bibfield  {journal} {\bibinfo  {journal} {Rev.
  Mod. Phys.}\ }\textbf {\bibinfo {volume} {84}},\ \bibinfo {pages} {1419}
  (\bibinfo {year} {2012})}\BibitemShut {NoStop}%
\bibitem [{\citenamefont {Simoncelli}\ \emph {et~al.}(2019)\citenamefont
  {Simoncelli}, \citenamefont {Marzari},\ and\ \citenamefont
  {Mauri}}]{simoncelli2019}%
  \BibitemOpen
  \bibfield  {author} {\bibinfo {author} {\bibfnamefont {M.}~\bibnamefont
  {Simoncelli}}, \bibinfo {author} {\bibfnamefont {N.}~\bibnamefont {Marzari}},
  \ and\ \bibinfo {author} {\bibfnamefont {F.}~\bibnamefont {Mauri}},\ }\href
  {\doibase 10.1038/s41567-019-0520-x} {\bibfield  {journal} {\bibinfo
  {journal} {Nature Phys.}\ }\textbf {\bibinfo {volume} {15}},\ \bibinfo
  {pages} {809} (\bibinfo {year} {2019})}\BibitemShut {NoStop}%
\bibitem [{\citenamefont {Wigner}(1932)}]{wigner1932}%
  \BibitemOpen
  \bibfield  {author} {\bibinfo {author} {\bibfnamefont {E.}~\bibnamefont
  {Wigner}},\ }\href {\doibase 10.1103/PhysRev.40.749} {\bibfield  {journal}
  {\bibinfo  {journal} {Phys. Rev.}\ }\textbf {\bibinfo {volume} {40}},\
  \bibinfo {pages} {749} (\bibinfo {year} {1932})}\BibitemShut {NoStop}%
\bibitem [{\citenamefont {Moyal}(1949)}]{moyal1949}%
  \BibitemOpen
  \bibfield  {author} {\bibinfo {author} {\bibfnamefont {J.~E.}\ \bibnamefont
  {Moyal}},\ }\href {\doibase 10.1017/S0305004100000487} {\bibfield  {journal}
  {\bibinfo  {journal} {Math. Proc. Camb. Philos. Soc.}\ }\textbf {\bibinfo
  {volume} {45}},\ \bibinfo {pages} {99} (\bibinfo {year} {1949})}\BibitemShut
  {NoStop}%
\bibitem [{\citenamefont {Groenewold}(1946)}]{Groenewold1946}%
  \BibitemOpen
  \bibfield  {author} {\bibinfo {author} {\bibfnamefont {H.}~\bibnamefont
  {Groenewold}},\ }\href {\doibase
  https://doi.org/10.1016/S0031-8914(46)80059-4} {\bibfield  {journal}
  {\bibinfo  {journal} {Physica}\ }\textbf {\bibinfo {volume} {12}},\ \bibinfo
  {pages} {405 } (\bibinfo {year} {1946})}\BibitemShut {NoStop}%
\bibitem [{\citenamefont {Blount}(1962)}]{BLOUNT1962305}%
  \BibitemOpen
  \bibfield  {author} {\bibinfo {author} {\bibfnamefont {E.~I.}\ \bibnamefont
  {Blount}},\ }\href {\doibase https://doi.org/10.1016/S0081-1947(08)60459-2}
  {\bibfield  {journal} {\bibinfo  {journal} {Solid State Phys.}\ }\textbf
  {\bibinfo {volume} {13}},\ \bibinfo {pages} {305 } (\bibinfo {year}
  {1962})}\BibitemShut {NoStop}%
\bibitem [{\citenamefont {Iotti}\ \emph {et~al.}(2017)\citenamefont {Iotti},
  \citenamefont {Dolcini},\ and\ \citenamefont {Rossi}}]{PhysRevB.96.115420}%
  \BibitemOpen
  \bibfield  {author} {\bibinfo {author} {\bibfnamefont {R.~C.}\ \bibnamefont
  {Iotti}}, \bibinfo {author} {\bibfnamefont {F.}~\bibnamefont {Dolcini}}, \
  and\ \bibinfo {author} {\bibfnamefont {F.}~\bibnamefont {Rossi}},\ }\href
  {\doibase 10.1103/PhysRevB.96.115420} {\bibfield  {journal} {\bibinfo
  {journal} {Phys. Rev. B}\ }\textbf {\bibinfo {volume} {96}},\ \bibinfo
  {pages} {115420} (\bibinfo {year} {2017})}\BibitemShut {NoStop}%
\bibitem [{\citenamefont {Zhan}\ \emph {et~al.}(2016)\citenamefont {Zhan},
  \citenamefont {Colom{\'e}s},\ and\ \citenamefont {Oriols}}]{Zhan2016}%
  \BibitemOpen
  \bibfield  {author} {\bibinfo {author} {\bibfnamefont {Z.}~\bibnamefont
  {Zhan}}, \bibinfo {author} {\bibfnamefont {E.}~\bibnamefont {Colom{\'e}s}}, \
  and\ \bibinfo {author} {\bibfnamefont {X.}~\bibnamefont {Oriols}},\ }\href
  {\doibase 10.1007/s10825-016-0875-5} {\bibfield  {journal} {\bibinfo
  {journal} {J. Comput. Electron.}\ }\textbf {\bibinfo {volume} {15}},\
  \bibinfo {pages} {1206} (\bibinfo {year} {2016})}\BibitemShut {NoStop}%
\bibitem [{\citenamefont {Nedjalkov}\ \emph {et~al.}(2013)\citenamefont
  {Nedjalkov}, \citenamefont {Selberherr}, \citenamefont {Ferry}, \citenamefont
  {Vasileska}, \citenamefont {Dollfus}, \citenamefont {Querlioz}, \citenamefont
  {Dimov},\ and\ \citenamefont {Schwaha}}]{NEDJALKOV2013}%
  \BibitemOpen
  \bibfield  {author} {\bibinfo {author} {\bibfnamefont {M.}~\bibnamefont
  {Nedjalkov}}, \bibinfo {author} {\bibfnamefont {S.}~\bibnamefont
  {Selberherr}}, \bibinfo {author} {\bibfnamefont {D.}~\bibnamefont {Ferry}},
  \bibinfo {author} {\bibfnamefont {D.}~\bibnamefont {Vasileska}}, \bibinfo
  {author} {\bibfnamefont {P.}~\bibnamefont {Dollfus}}, \bibinfo {author}
  {\bibfnamefont {D.}~\bibnamefont {Querlioz}}, \bibinfo {author}
  {\bibfnamefont {I.}~\bibnamefont {Dimov}}, \ and\ \bibinfo {author}
  {\bibfnamefont {P.}~\bibnamefont {Schwaha}},\ }\href {\doibase
  https://doi.org/10.1016/j.aop.2012.10.001} {\bibfield  {journal} {\bibinfo
  {journal} {Ann. Phys.}\ }\textbf {\bibinfo {volume} {328}},\ \bibinfo {pages}
  {220 } (\bibinfo {year} {2013})}\BibitemShut {NoStop}%
\bibitem [{\citenamefont {Nedjalkov}\ \emph {et~al.}(2011)\citenamefont
  {Nedjalkov}, \citenamefont {Querlioz}, \citenamefont {Dollfus},\ and\
  \citenamefont {Kosina}}]{Nedjalkov2011}%
  \BibitemOpen
  \bibfield  {author} {\bibinfo {author} {\bibfnamefont {M.}~\bibnamefont
  {Nedjalkov}}, \bibinfo {author} {\bibfnamefont {D.}~\bibnamefont {Querlioz}},
  \bibinfo {author} {\bibfnamefont {P.}~\bibnamefont {Dollfus}}, \ and\
  \bibinfo {author} {\bibfnamefont {H.}~\bibnamefont {Kosina}},\ }\enquote
  {\bibinfo {title} {Wigner function approach},}\ in\ \href {\doibase
  10.1007/978-1-4419-8840-9_5} {\emph {\bibinfo {booktitle} {Nano-Electronic
  Devices: Semiclassical and Quantum Transport Modeling}}},\ \bibinfo {editor}
  {edited by\ \bibinfo {editor} {\bibfnamefont {D.}~\bibnamefont {Vasileska}}\
  and\ \bibinfo {editor} {\bibfnamefont {S.~M.}\ \bibnamefont {Goodnick}}}\
  (\bibinfo  {publisher} {Springer New York},\ \bibinfo {address} {New York,
  NY},\ \bibinfo {year} {2011})\ pp.\ \bibinfo {pages} {289--358}\BibitemShut
  {NoStop}%
\bibitem [{\citenamefont {Gebauer}\ and\ \citenamefont
  {Car}(2004)}]{PhysRevB.70.125324}%
  \BibitemOpen
  \bibfield  {author} {\bibinfo {author} {\bibfnamefont {R.}~\bibnamefont
  {Gebauer}}\ and\ \bibinfo {author} {\bibfnamefont {R.}~\bibnamefont {Car}},\
  }\href {\doibase 10.1103/PhysRevB.70.125324} {\bibfield  {journal} {\bibinfo
  {journal} {Phys. Rev. B}\ }\textbf {\bibinfo {volume} {70}},\ \bibinfo
  {pages} {125324} (\bibinfo {year} {2004})}\BibitemShut {NoStop}%
\bibitem [{\citenamefont {Frensley}(1990)}]{RevModPhys.62.745}%
  \BibitemOpen
  \bibfield  {author} {\bibinfo {author} {\bibfnamefont {W.~R.}\ \bibnamefont
  {Frensley}},\ }\href {\doibase 10.1103/RevModPhys.62.745} {\bibfield
  {journal} {\bibinfo  {journal} {Rev. Mod. Phys.}\ }\textbf {\bibinfo {volume}
  {62}},\ \bibinfo {pages} {745} (\bibinfo {year} {1990})}\BibitemShut
  {NoStop}%
\bibitem [{\citenamefont {Ziman}(1960)}]{ziman1960electrons}%
  \BibitemOpen
  \bibfield  {author} {\bibinfo {author} {\bibfnamefont {J.~M.}\ \bibnamefont
  {Ziman}},\ }\href@noop {} {\emph {\bibinfo {title} {{Electrons and phonons:
  the theory of transport phenomena in solids}}}}\ (\bibinfo  {publisher}
  {Oxford university press},\ \bibinfo {year} {1960})\BibitemShut {NoStop}%
\bibitem [{\citenamefont {Verdi}\ and\ \citenamefont
  {Giustino}(2015)}]{PhysRevLett.115.176401}%
  \BibitemOpen
  \bibfield  {author} {\bibinfo {author} {\bibfnamefont {C.}~\bibnamefont
  {Verdi}}\ and\ \bibinfo {author} {\bibfnamefont {F.}~\bibnamefont
  {Giustino}},\ }\href {\doibase 10.1103/PhysRevLett.115.176401} {\bibfield
  {journal} {\bibinfo  {journal} {Phys. Rev. Lett.}\ }\textbf {\bibinfo
  {volume} {115}},\ \bibinfo {pages} {176401} (\bibinfo {year}
  {2015})}\BibitemShut {NoStop}%
\bibitem [{\citenamefont {Monserrat}\ and\ \citenamefont
  {Vanderbilt}(2016)}]{PhysRevLett.117.226801}%
  \BibitemOpen
  \bibfield  {author} {\bibinfo {author} {\bibfnamefont {B.}~\bibnamefont
  {Monserrat}}\ and\ \bibinfo {author} {\bibfnamefont {D.}~\bibnamefont
  {Vanderbilt}},\ }\href {\doibase 10.1103/PhysRevLett.117.226801} {\bibfield
  {journal} {\bibinfo  {journal} {Phys. Rev. Lett.}\ }\textbf {\bibinfo
  {volume} {117}},\ \bibinfo {pages} {226801} (\bibinfo {year}
  {2016})}\BibitemShut {NoStop}%
\bibitem [{\citenamefont {Cao}\ \emph {et~al.}(2020)\citenamefont {Cao},
  \citenamefont {Querales-Flores}, \citenamefont {Fahy},\ and\ \citenamefont
  {Savi\'c}}]{CAO2020100172}%
  \BibitemOpen
  \bibfield  {author} {\bibinfo {author} {\bibfnamefont {J.}~\bibnamefont
  {Cao}}, \bibinfo {author} {\bibfnamefont {J.}~\bibnamefont
  {Querales-Flores}}, \bibinfo {author} {\bibfnamefont {S.}~\bibnamefont
  {Fahy}}, \ and\ \bibinfo {author} {\bibfnamefont {I.}~\bibnamefont
  {Savi\'c}},\ }\href {\doibase https://doi.org/10.1016/j.mtphys.2019.100172}
  {\bibfield  {journal} {\bibinfo  {journal} {Materials Today Physics}\
  }\textbf {\bibinfo {volume} {12}},\ \bibinfo {pages} {100172} (\bibinfo
  {year} {2020})}\BibitemShut {NoStop}%
\bibitem [{\citenamefont {Jiao}\ \emph {et~al.}(2012)\citenamefont {Jiao},
  \citenamefont {Jiang}, \citenamefont {Feng}, \citenamefont {Xu},
  \citenamefont {Cao}, \citenamefont {Xu}, \citenamefont {Feng}, \citenamefont
  {Yamada}, \citenamefont {Matsubayashi},\ and\ \citenamefont
  {Uwatoko}}]{doi:10.1063/1.4727957}%
  \BibitemOpen
  \bibfield  {author} {\bibinfo {author} {\bibfnamefont {W.~H.}\ \bibnamefont
  {Jiao}}, \bibinfo {author} {\bibfnamefont {S.}~\bibnamefont {Jiang}},
  \bibinfo {author} {\bibfnamefont {C.~M.}\ \bibnamefont {Feng}}, \bibinfo
  {author} {\bibfnamefont {Z.~A.}\ \bibnamefont {Xu}}, \bibinfo {author}
  {\bibfnamefont {G.~H.}\ \bibnamefont {Cao}}, \bibinfo {author} {\bibfnamefont
  {M.}~\bibnamefont {Xu}}, \bibinfo {author} {\bibfnamefont {D.~L.}\
  \bibnamefont {Feng}}, \bibinfo {author} {\bibfnamefont {A.}~\bibnamefont
  {Yamada}}, \bibinfo {author} {\bibfnamefont {K.}~\bibnamefont
  {Matsubayashi}}, \ and\ \bibinfo {author} {\bibfnamefont {Y.}~\bibnamefont
  {Uwatoko}},\ }\href {\doibase 10.1063/1.4727957} {\bibfield  {journal}
  {\bibinfo  {journal} {AIP Advances}\ }\textbf {\bibinfo {volume} {2}},\
  \bibinfo {pages} {022148} (\bibinfo {year} {2012})},\ \Eprint
  {http://arxiv.org/abs/https://doi.org/10.1063/1.4727957}
  {https://doi.org/10.1063/1.4727957} \BibitemShut {NoStop}%
\bibitem [{\citenamefont {Ren}\ \emph {et~al.}(2011)\citenamefont {Ren},
  \citenamefont {Taskin}, \citenamefont {Sasaki}, \citenamefont {Segawa},\ and\
  \citenamefont {Ando}}]{PhysRevB.84.075316}%
  \BibitemOpen
  \bibfield  {author} {\bibinfo {author} {\bibfnamefont {Z.}~\bibnamefont
  {Ren}}, \bibinfo {author} {\bibfnamefont {A.~A.}\ \bibnamefont {Taskin}},
  \bibinfo {author} {\bibfnamefont {S.}~\bibnamefont {Sasaki}}, \bibinfo
  {author} {\bibfnamefont {K.}~\bibnamefont {Segawa}}, \ and\ \bibinfo {author}
  {\bibfnamefont {Y.}~\bibnamefont {Ando}},\ }\href {\doibase
  10.1103/PhysRevB.84.075316} {\bibfield  {journal} {\bibinfo  {journal} {Phys.
  Rev. B}\ }\textbf {\bibinfo {volume} {84}},\ \bibinfo {pages} {075316}
  (\bibinfo {year} {2011})}\BibitemShut {NoStop}%
\bibitem [{\citenamefont {Analytis}\ \emph {et~al.}(2010)\citenamefont
  {Analytis}, \citenamefont {McDonald}, \citenamefont {Riggs}, \citenamefont
  {Chu}, \citenamefont {Boebinger},\ and\ \citenamefont
  {Fisher}}]{natphys.bi2se3}%
  \BibitemOpen
  \bibfield  {author} {\bibinfo {author} {\bibfnamefont {J.~G.}\ \bibnamefont
  {Analytis}}, \bibinfo {author} {\bibfnamefont {R.~D.}\ \bibnamefont
  {McDonald}}, \bibinfo {author} {\bibfnamefont {S.~C.}\ \bibnamefont {Riggs}},
  \bibinfo {author} {\bibfnamefont {J.-H.}\ \bibnamefont {Chu}}, \bibinfo
  {author} {\bibfnamefont {G.~S.}\ \bibnamefont {Boebinger}}, \ and\ \bibinfo
  {author} {\bibfnamefont {I.~R.}\ \bibnamefont {Fisher}},\ }\href {\doibase
  10.1038/nphys1861} {\bibfield  {journal} {\bibinfo  {journal} {Nature Phys.}\
  }\textbf {\bibinfo {volume} {6}},\ \bibinfo {pages} {960} (\bibinfo {year}
  {2010})}\BibitemShut {NoStop}%
\end{thebibliography}


\end{document}